\documentclass[lettersize,journal]{IEEEtran}
\pdfoutput=1
\usepackage{amsmath,amsfonts}
\usepackage{algorithmic}
\usepackage{algorithm}
\usepackage{array}
\usepackage[caption=false,font=normalsize,labelfont=sf,textfont=sf]{subfig}
\usepackage{textcomp}
\usepackage{stfloats}
\usepackage{url}
\usepackage{verbatim}
\usepackage{graphicx}
\usepackage{cite}
\usepackage{hyperref}
\hypersetup{hidelinks,
	colorlinks=true,
	allcolors=black,
	pdfstartview=Fit,
	breaklinks=true}

\usepackage{stfloats}
\usepackage{amsmath}
\usepackage{booktabs}
\usepackage{comment}
\usepackage{ragged2e}
\usepackage[caption=false, font=normalsize, labelfont=sf, textfont=sf]{subfig}
\UseRawInputEncoding
\begin{document}
\title{Energy-Efficient Blockchain-enabled User-Centric Mobile Edge Computing}
\author{Langtian Qin, Hancheng Lu,~\IEEEmembership{Senior Member,~IEEE}, Yuang Chen, Zhuojia Gu, Dan Zhao, \\and Feng Wu,~\IEEEmembership{Fellow,~IEEE}
\IEEEcompsocitemizethanks{\IEEEcompsocthanksitem L.Qin, H.Lu, Y.Chen, Z.Gu, D.Zhao and F.Wu are with the Department of Electronic Engineering and Information Science, University of Science and Technology of
China, Hefei 230027, China.  \protect\\
E-mail: qlt315@mail.ustc.edu.cn; hclu@ustc.edu.cn; \{yuangchen21, guzj, zd2019@mail.ustc.edu.cn\};      fengwu@ustc.edu.cn
 }
}

\maketitle
\begin{abstract}
In the traditional mobile edge computing (MEC) system, the availability of MEC services is greatly limited for the edge users of the cell due to serious signal attenuation and inter-cell interference. User-centric MEC (UC-MEC) can be seen as a promising solution to address this issue. In UC-MEC, each user is served by a dedicated access point (AP) cluster enabled with MEC capability instead of a single MEC server, however, at the expense of more energy consumption and greater privacy risks. To achieve efficient and reliable resource utilization with user-centric services, we propose an energy efficient blockchain-enabled UC-MEC system where blockchain operations and resource optimization are jointly performed. Firstly, we design a resource-aware, reliable, replicated, redundant, and fault-tolerant (R-RAFT) consensus mechanism to implement secure and reliable resource trading. Then, an optimization framework based on alternating direction method of multipliers (ADMM) is proposed to minimize the total energy consumed by wireless transmission, consensus and task computing, where APs clustering, computing resource allocation and bandwidth allocation are jointly considered. Simulation results show superiority of the proposed UC-MEC system over reference schemes, at most $33.96\%$ reduction in the total delay and $48.77\%$ reduction in the total energy consumption.
\end{abstract}


\section{Introduction}
\IEEEPARstart {W}{ITH} the rapid development of mobile Internet, the growing variety of mobile applications put forward higher requirements for resource capability of smart devices\cite{Luo21}. By moving the services and functions originally located in the cloud to the user side, mobile edge computing (MEC) provides powerful computing, storage, networking and communication capabilities at the edge of wireless access networks. {\color{black}However, in traditional cellular-based MEC networks that adopted by most of the existing works about MEC, each user will only be provided with transmission and computing services by a single BS integrated with MEC servers. Users at the edge of the cell will prone to suffer serious signal attenuation and inter-cell interference, which may significantly increase the transmission delay or even cause the offloading failure. Moreover, resources of a single cell may not be able to meet the requirements of users with resource-consumption mobile applications such as augmented reality (AR), virtual reality (VR), et al.}

{\color{black}As an emerging technology of 5G and beyond, user-centric network (UCN) can be seen as a reliable solution to overcome the above problems\cite{Pan2018}. In UCN, each user will be served by a dynamically divided AP set, which is called AP cluster\cite{Caso2021}. Each AP cluster can be divided adaptively according to the location and network condition of users to provide seamless wireless transmission service. By integrating MEC servers in the AP, user-centric MEC (UC-MEC)\cite{Qin2023} breaks the concept of ``cell'' in traditional cellular-based MEC, and can further expand the computing and communication resources for task offloading in MEC. In UC-MEC, When a user requests MEC services, it will be collaboratively served by a dynamically divided set of APs, i.e., AP cluster. Each AP in the cluster processes part of the offloaded tasks of the user. Through UC-MEC, users can be provided with efficient and reliable wireless transmission and task processing services wherever they are.}

{\color{black}In spite of the overarching merits, user-centric service mode will make the issues of security and privacy more prominent. Firstly, users need to transmit signals to multiple APs instead of a single AP through wireless channels. In addition, since it is very difficult to connect each AP with ideal wired backhaul links, APs generally need to transmit information through wireless backhaul to ensure cooperative services\cite{Chen2016}. Frequent wireless transmissions increase the risk of being attacked compared with traditional wireless networks due to the openness of wireless channels. Secondly, since the AP cluster will change dynamically with the user location and network condition, the access and authorization management of the AP cluster will become more important. If any AP in the AP cluster is hijacked, the user's privacy information may be easily disclosed during the offloading process. However, the traditional centralized security solutions has the hidden danger of single point of failure. When the network controller fails or is maliciously hijacked, the security and privacy of users in the network will be difficult to protect. Also, the centrality of the controller increases communication overhead and network latency\cite{Boateng2022}.}

{\color{black}Blockchain can be viewed as an promising technology to solve these problems in UC-MEC due to its immutability, decentralization, transparency, security, and privacy features. In blockchain-enabled UC-MEC system, each node maintains an identical distributed ledger, which records network resource trading information between users and APs. The ledger is totally transparent and tamper-free. After the offloading  completed, the network node executes the consensus mechanism, and packages the transaction information into blocks to record on the blockchain, which can prevent data loss and tampering caused by node failure. In addition, the advent of the blockchain-based smart contract\cite{Zou2021} can establish the access control and authorization mechanism of the AP cluster in UC-MEC conveniently and efficiently.}

{\color{black}To exploit the benefits of blockchain-enabled UC-MEC, there still remain the following research gaps to be addressed. Firstly, in UC-MEC systems, multiple APs are dynamically formed into an AP cluster to realize a user-centric service. It means that all APs are potential to allocate transmission and MEC resources for users. In this case, a decentralized mechanism is required to ensure secure and reliable resource trading between APs and users. Secondly, blockchain-enabled UC-MEC can significantly improve the quality of service (QoS) and ensure the security and privacy of users, however, at the expense of surge in system energy consumption \cite{Pan2018}. To minimize the total energy consumption of blockchain-enabled UC-MEC, it is necessary to jointly consider the energy consumption during task offloading and consensus process. These two issues are coupled with APs clustering decision and resource allocation in blockchain-enabled UC-MEC systems, which are more challenging than that in traditional blockchain-enabled MEC systems.}

\subsection{Related Work}
Blockchain has been viewed as a promising technology to enhance security and reliability of MEC systems in a distributed manner \cite{Fan2021}. There exist many studies on the combination of blockchain and MEC. Some studies considered the consensus process as a task offloading to the MEC server to reduce consensus delay \cite{Wangsi2021, Xiongze2018, Liu2018}. The authors of \cite{Wangsi2021} used consortium blockchain and smart contract to achieve trusted resource allocation in edge computing of vehicle network. An edge computing resource management framework based on optimal pricing is proposed in \cite{Xiongze2018}, where the mining process can be offloaded to the edge computing service provider. In \cite{Liu2018}, authors considered that mining tasks can be offloaded to nearby edge computing nodes and encrypted hashes of blocks can be cached in MEC servers. There also exist some studies on consideration of both computing task offloading and block mining in the consensus process \cite{Nguyen, Feng2020}. In \cite{Nguyen}, authors proposed a new collaborative task offloading and block mining algorithm for blockchain-based MEC systems. In \cite{Feng2020}, joint optimization was performed on offloading decisions, power allocation, block size, and block interval to maximize the computing speed of the MEC system and the transaction throughput of the blockchain system. With the rise of machine learning in MEC systems \cite{Hu2021}, blcokchain was used to ensure the privacy of distributed learning \cite{Fan2021}. However, most of these studies considered resource allocation in task offloading and blockchain operations separately, which might lead to low resource utilization efficiency. Moreover, existing studies focused on blockchain operations in a server-centric manner. For UC-MEC where all APs with MEC capability are potential service providers for users, new consensus processes involving all APs should be designed.

Energy consumption has also been focused on in traditional MEC systems. {\color{black}In \cite{Apostolopoulos2020}, the authors considered users’risk-seeking or loss-aversion behavior in their offloading process. A non-cooperative game among the users is formulated and  a distributed low-complexity algorithm is proposed to obtain the corresponding Pure Nash Equilibrium (PNE), i.e., optimal data offloading strategy. By adopting the satisfaction games and approximate computing, the authors of \cite{Irtija2022} introduced an energy efficient solution in MEC-enabled fully autonomous aerial systems (FAAS) to obtain the optimal partial offloading decisions under minimum QoS prerequisites. The authors in \cite{Bishoyi2021}  modeled joint cost and energy-efficient task offloading in the MEC-enabled healthcare system by Stackelberg game. The optimal task offloading decision can be derived in a distributed manner by using alternating direction method of multipliers (ADMM) method. In \cite{Bishoyi2022}, the authors modeled the economic interaction between the MEC server and users using the Nash bargaining theory to minimizes the energy consumption of the MEC server without compromising on the quality-of-experience (QoE) of the users. The authors of \cite{Du2022} proposed a software-defined networking (SDN) based framework for edge and cloud computing system.  An evolutionary stackelberg differential game based dynamic pricing and computing resource allocation mechanism is proposed to optimize computing resource utilization and satisfy the time-varying computational tasks. In \cite{Bozorgchenani2021}, the task offloading in MEC is modeled as a constrained multi-objective optimization problem (CMOP). The authors proposed an evolutionary algorithm to find the best trade-offs between energy consumption and task processing delay. However, in these studies, the energy consumption for wireless transmission has not been well considered, which is nontrivial in UC-MEC systems supporting multiple wireless connections for users. Additionally, APs in UC-MEC are connected via wireless links to ensure the flexibility of deployment\cite{Guo2020}. In this case, the energy consumption for wireless communication between APs cannot be ignored.

}

\subsection{Contributions}
To address aforementioned issues, we propose an energy-efficient blockchain-enabled UC-MEC system. Different from traditional MEC, in the proposed system, we provide users with user-centric computing services, and then perform joint optimization of APs clustering decision, computing resource and bandwidth allocation to achieve secure, reliable and efficient resource utilization. The main contributions of this paper are summarized as follows.
\begin{itemize}

\item{To implement secure and reliable resource trading in UC-MEC, we design a resource-aware consensus mechanism based on replicated, redundant, and fault-tolerant (RAFT) mechanism, which is called R-RAFT. In R-RAFT, we propose the resource-aware \emph{Leader} election mechanism to replace the random selection of \emph{Leader} in traditional RAFT. With R-RAFT, the efficiency of the consensus process is significantly improved.}

\item{We analyze the energy consumption of the proposed blockchain-enabled UC-MEC. The energy consumption for wireless transmission and consensus process is jointly considered with that for task computing. Based on our analytical work, we formulate the energy consumption minimization problem by jointly optimizing APs clustering, computing resource allocation and bandwidth allocation.}

\item{To solve the formulated problem, we propose a parallel optimization framework based on alternating direction method of multipliers (ADMM). Particularly, we firstly decompose the problem into some sub-problems. Then we convert the sub-problem into a D.C programming problem and propose a sub-gradients iterative algorithm to solve the converted problem in parallel by all APs.}
\end{itemize}

The simulation results show that the proposed ADMM-based scheme outperforms reference schemes in terms of total delay and energy consumption. Compared with traditional MEC, the proposed blockchain-enabled UC-MEC system can reduce the total delay .

\subsection{Organization}
The rest of the paper is organized as follows. The blockchain-enabled UC-MEC system is modeled in Section II. In Section III, the energy consumption of the proposed UC-MEC system is analyzed and then the energy consumption minimization problem is formulated. To solve the formulated problem, joint APs clustering and resource allocation optimization based on ADMM is performed in Section IV. Simulation results are present in Section V. Finally, conclusion and future works are given in Section VI.

\begin{figure}[tbp]
	\setlength{\belowcaptionskip}{-0.4cm}
	\centering
	\includegraphics[width=0.5\textwidth]{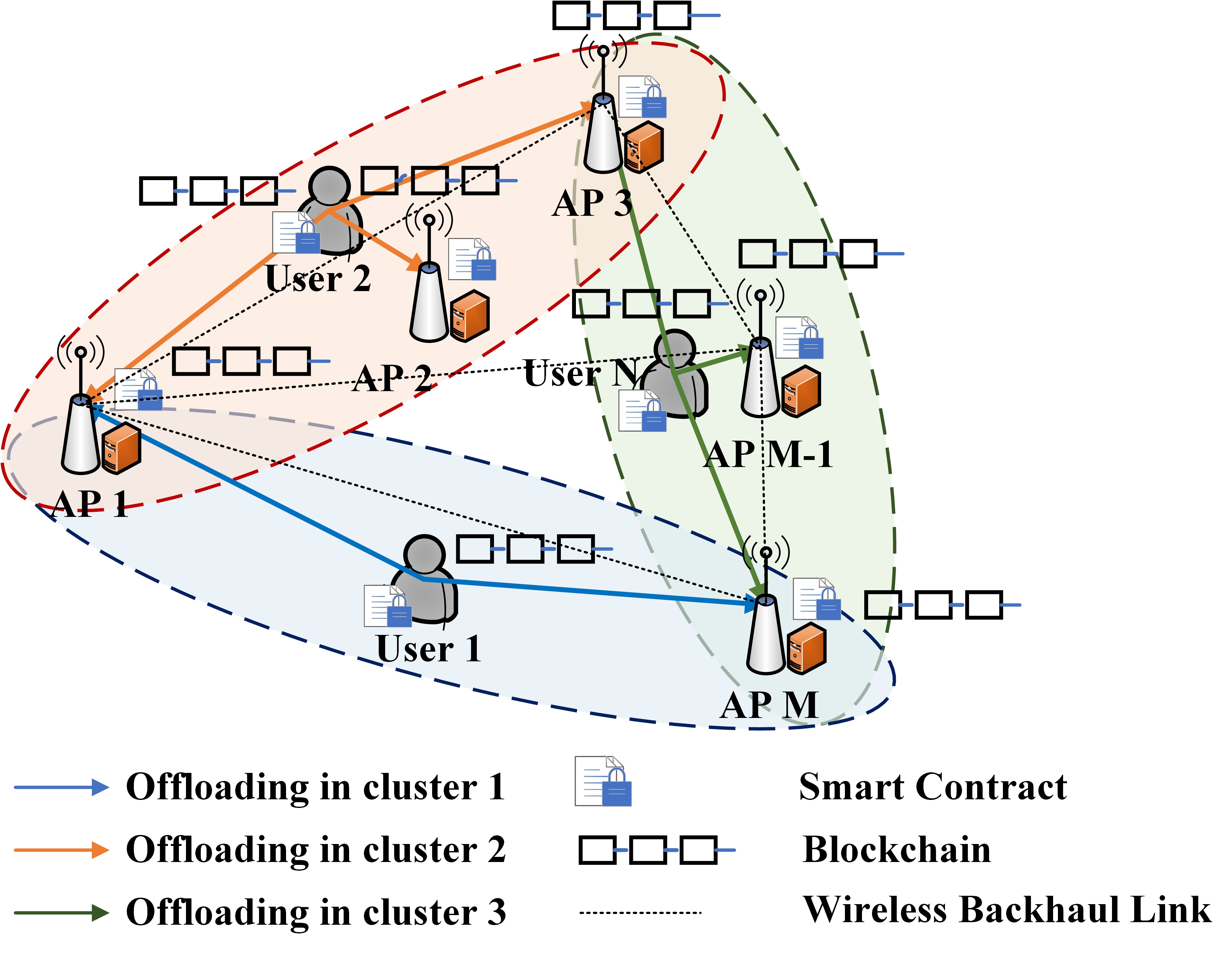}
	\caption{The framework of blockchain-enabled UC-MEC.}
	\label{multi}
\end{figure}

\section{System Model}
In this section, we first introduce the UC-MEC framework, including the network and the task offloading model of UCMEC. Then we describe the blockchain based resource trading process in UC-MEC, including our proposed R-RAFT consensus mechanism. Some key notations are shown in Table I.

\subsection{Network Model}
\par As shown in Fig. 1,  we consider a blockchain-enabled UC-MEC system with $M$ APs and $N$ users. There are two types of wireless links in UC-MEC, i.e., the wireless data transmission links between users and APs, and the wireless backhaul links among APs.
We assume that the set of APs is represented as $\mathcal{M} = \left \{1, 2, 3,\cdots, M\right \}$. Each AP is equipped with $X>N$ antennas and is integrated with a MEC server which contains certain bandwidth and computing capability. Considering the heterogeneity of APs, the bandwidth and the computing capability of each AP are different, which are indicated by $B_{m}$ (in MHz) and $C_{m}$ (in CPU-cycle frequency), respectively. We denote the set of users in the network as $\mathcal{N} = \left \{ 1,2,\cdots,N\right \}$, and each user will be served by a specific AP cluster (i.e., a collection of several APs).

\subsection{Task Offloading Model}
\par In a certain time slot, each user generates a task that need to offload to the MEC-enabled APs. We use a triple $Task_n = \{L_n,\rho_n, D_n^{t}\}$ to express the task that user $n$ ($n\in\mathcal{N}$) needs to offload, where $L_n$, $\rho_n$ and $D_n^{t}$ stand for the task data size (in bit), computing density (in CPU cycles per bit) and maximum tolerated delay (in second) \cite{Lei}. When user $n$ generates a task and sends the offloading request to the network, APs in the network will decide whether to join the cluster of user $n$ or not. Similar to \cite{Bo2021}, we use a continuous variable $a_{m,n}\in\left [0,1\right]$ to indicate the cluster decision of AP $m$ for user $n$. If $a_{m,n}=0$, AP $m$ will not join the AP cluster of user $n$; if $a_{m,n}\in \left( 0,1 \right ]$, AP $m$ will join the AP cluster of user $n$, and responsible for the corresponding proportion of offloading task. We define $\mit\Phi_n$ as the AP cluster of user $n$ and the users served by $\mit\Phi_n$ is indicated by $\mit\Omega_{n}$. The users who share the same AP cluster with user $n$ are called intra-cluster users, while the remaining users are called inter-cluster users. In this case, the total signal received by AP $m$ when providing edge services for user $n$ includes the useful signal of user $n$ and the interference signals of inter-cluster users and intra-cluster users, which is calculated as follows
\begin{equation}\label{signal}
	\begin{aligned}
		\boldsymbol s_{m,n}=&\sqrt{p_n^{u}} \boldsymbol g_{m,n} x_n+\!\!\!\!\!
		\sum_{\tiny{\begin{array}{c}
					v\!\neq \!n,\\
					v\!\in \!\mit\Omega_n\end{array}}}\!\!\!\!\sqrt{p_v^u} \boldsymbol g_{m,v}x_v\\
		&+\!\!\!\sum_{\tiny{\begin{array}{c}
					w\!\neq \!n,\\
					w\!\notin \!\mit\Omega_n\end{array}}}\!\!\!\!\sqrt{p_w^u}\boldsymbol g_{m,w}x_w+z,
	\end{aligned}
\end{equation}
where $p_{n}^u$ is the transmit signal power of user $n$, $\boldsymbol g_{m,n}\in \mathbb{C}^{X\times1}$ is the complex channel coefficient between user $n$ and AP $m$, $z$ is the Gaussian white noise with mean 0 and variance $\sigma^{2}$, $x_n \sim \mathcal{CN}(0,1) $ is the complex signal send by user $n$.
\par Through beamforming technology, the interference of intra-cluster users can be completely eliminated\cite{zhu2018}. The beamforming vector of user $n$ be calculated as
\begin{equation}
	{\boldsymbol w}_n=\frac{({\boldsymbol I}_{A\left|\mit\Phi_n\right|}-{\boldsymbol G}_{-n}\boldsymbol G_{-n}^\dagger)\boldsymbol g_n^n}{{\left\|({\boldsymbol I}_{A\left|\mit\Phi_n\right|}-{\boldsymbol G}_{-n}\boldsymbol G_{-n}^\dagger)\boldsymbol g_n^n\right\|}_2},
\end{equation}
where $\boldsymbol g^n_n=[\cdots, \boldsymbol g_{m,n}, \cdots]^{T}_{m\in \mit\Phi_n}$, $ \boldsymbol G_{-n}=[\cdots,  (\boldsymbol g_v^n),\cdots]_{v\neq n,v\in \mit\Omega_n}$ and $\boldsymbol g^n_v=[\cdots, \boldsymbol g_{m,v}, \cdots]^{T}_{m\in \mit\Phi_n}$.

\par Assume the computing resources and the bandwidth AP $m$ allocates to user $n$ are $c_{m,n}$ and $b_{m,n}$, respectively. Both $c_{m,n}$ and $b_{m,n}$ are continuous variables, and cannot exceed the computing capability and bandwidth of AP $m$.  AP $m$ will not allocate resources to the user who are not within the service range. To allocate bandwidth and ensure that users can send signals to different APs in the AP cluster, the orthogonal frequency division multiple access (OFDMA) method is adopt in this paper \cite{Wang2020}. When AP $m$ provides bandwidth for user $n$ to transmit task data, the uplink transmission rate of user $n$ can be expressed by

\begin{equation}
    r_{m,n}=b_{m,n}\mathrm{log_2}(1 + \frac{p_n^{u}\left|\boldsymbol {w}_n^H\boldsymbol{g}^n_n\right|^2}{\sum\limits_{w\notin \mit\Omega_n} p_w^u\left| \boldsymbol{w}_n^H   \boldsymbol{g}^n_w\right|^2 +\left| \boldsymbol w_n\right|^2\sigma^2}).
\end{equation}

\par After the AP cluster finishes processing the user's task, the results will be integrated and return to the user. After verifying the returned data, the user will pay the AP cluster remuneration according to the amount of resources provided by the AP cluster. Since the size of downlink data is relatively small, we ignore the overhead of downlink data in this paper.

 \begin{figure*}[htbp]
	\setlength{\belowcaptionskip}{-0.4cm}
	\centering
	\includegraphics[width=1\textwidth]{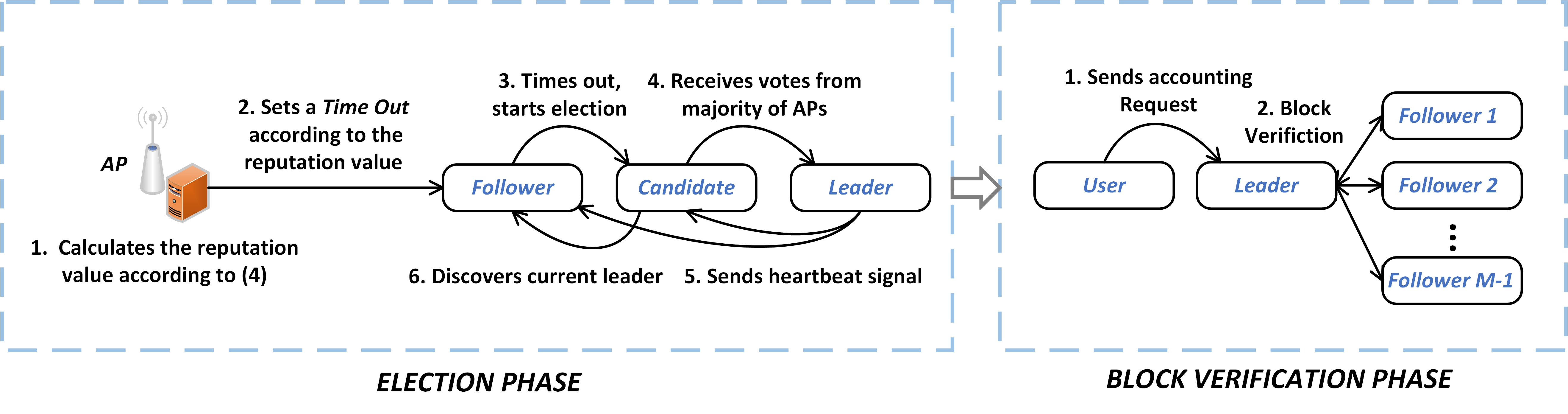}
	\caption{Illustration of R-RAFT consensus mechanism}
	\label{multi}
\end{figure*}

\subsection{Blockchain-based Consensus Model}

\par To establish a multi-party trusted resource transaction environment and ensure the data safety and privacy of users, blockchain is deployed in our UC-MEC system. Due to the limited computing and storage capability of users, we assume users as the light weight nodes, who are responsible for generating transactions and requests, and only keep part of the ledger. As the full nodes, APs own the complete ledger and are responsible for the consensus process. After the resource trading is completed, the blockchain system will record the trading information (i.e., timestamp, AP ID, user ID, amount of computing resources and bandwidth provided by the AP cluster, amount of remuneration paid by user, information of computing task, et.al.\cite{Shi2022}) by executing the consensus mechanism. Furthermore, a smart contract with a series of security mechanisms is deployed both on the user side and AP side to manage the process of APs clustering, resource trading and consensus process, which can ensure the credibility of resource trading and the data security of users.  Next, we will introduce RAFT and our R-RAFT consensus mechanism.

\par RAFT is an algorithm implements in distributed system for reaching consensus among nodes, which has been widely used in Internet of Things (IoT) due to its simplicity of process and high speed of consensus\cite{Hou202}. In RAFT, all nodes are divided into three roles:1) \emph{Follower}; 2) \emph{Candidate}; and 3) \emph{Leader}. The process of RAFT consists of two phases, i.e., the election phase and the block verification phase. In the election phase, all nodes become \emph{Followers} by default and decide to elect the \emph{Leader}. RAFT uses a heartbeat mechanism to trigger the \emph{Leader} election. At first, each node will randomly initialize a \emph{time out} value, and if a node does not receive the heartbeat signal from the \emph{Leader} before the end of the \emph{time out}, it will convert to a \emph{Candidate}.  Each \emph{Candidate} will broadcast to all other nodes for canvassing. After receiving the canvass request, each \emph{Follower}  will vote for the \emph{Candidate} who send the request, and the \emph{Candidate} with the largest number of votes will become the \emph{Leader}. After the \emph{Leader} election, all other non-\emph{Leader} nodes will become \emph{Followers} and assist the \emph{Leader} to verify the block. In the block verification phase, the user first sends an accounting request to the \emph{Leader}. Then the \emph{Leader} generates and copies the block to other \emph{Followers}. After that, Other \emph{Followers} verify the block and return the confirmation signal to the \emph{Leader}.

 As described above, since the \emph{Leader} is responsible for the generation and duplicate of the block, it needs to consume much more resources compared to other \emph{Followers}, thus the \emph{Leader} should be selected carefully. However, the \emph{Leader} is randomly assigned in RAFT which is far from optimal. Besides, RAFT is prone to multiple \emph{Candidates} with the same number of votes during the election phase, which will lead to the failure of the election, and all the nodes need to wait for a \emph{time out} long and restart the election process again, thus greatly increase the consensus overhead. Therefore, we propose a resource-aware RAFT algorithm named R-RAFT to improve the performance of the original RAFT. The illustration of R-RAFT is shown in Fig. 2.

\par In R-RAFT, after APs finish processing the computing tasks, the remaining computing resources and bandwidth are denoted as $\overline{C}_{m}$ and $\overline{B}_{m}$, respectively. During the consensus, the \emph{Leader} and \emph{Followers} need to send state confirmation signals (such as heartbeat packets, vote requests, confirm replies, et.al.) and blocks via the wireless channel. To simplify the communication model of R-RAFT, we assume the data size of state confirmation as $L^s$. Let $L^b$ and $D^i$ denote the size and the generation interval of blocks, respectively. The bandwidth and the computing resources that the \emph{Leader} needs to consume in the consensus are far more than those of \emph{Followers}. Therefore, we can choose the AP with the most abundant resources as the \emph{Leader}. We use a variable called reputation to describe the communication and the computing capability of APs in the consensus process. Assume all remaining resources of APs are used for consensus, we can define the reputation of AP ${m}$ as
\begin{equation}
    R_m =  \frac{\overline{C}_m}{L^b} + \frac{\overline{B}_m}{L^s+L^b}.
\end{equation}
The larger amount of remaining resources means that the communication and the computing capability of an AP in the consensus process will be stronger, and its reputation will be higher.
\par To prevent the uncertainty of \emph{Leader} election in the original RAFT, we propose a reputation-based \emph{time out} initialization strategy where \emph{time out} is proportional to $\frac{1}{R_m}$. The node with shorter \emph{time out} can become \emph{Candidate} faster and have a first-mover advantage of becoming \emph{Leader}.  In this paper,  we use a probability-based model to describe the behavior of APs in the R-RAFT mechanism. We define the probability of AP $m$ becoming the \emph{Leader} in the election as $P(m = \emph{Leader})$, $\forall m \in \mathcal{M}$ which is positively correlated with $R_{m}$ as
\begin{equation}
    P(m = \emph{Leader})=\frac{R_m}{R},
\end{equation}
where $R$ is the maximum reputation value of all APs, which can be calculated by
\begin{equation}
  R = \max_{m \in \mathcal{M}} \left \{ R_{m}\right\}.
\end{equation}
\par Since all APs will become a \emph{Leader} or a \emph{Follower} after the election, thus $P(m = \emph{Follower}) = 1 - P(m=\emph{Leader})$. Next, we model the transmission process among the \emph{Leader} and the \emph{Followers}. Assume the backhaul channel among APs and the access channel between APs and users adopt different spectrum. Therefore, an AP only needs to consider the interference of other APs. Suppose $p_m^{a}$ is the signal power sent by AP $m$, $p_m^{i}$ is the interference signal power of other APs, and $h_{m_1,m_2}$ is the channel coefficient between AP $m_1$ and AP $m_2$. $h_{{m_1},{m_2}}$ can be generated by $h_{m_1,m_2} = \frac {h}{\sqrt{d_{m_1,m_2}^{\gamma}}}$, $h \sim \mathcal{CN}(0,1)$ , where $\gamma$ is the path loss, $d_{m_1,m_2}$ is the distance (in kilometers) between AP ${m_1}$ and AP ${m_2}$. Similar to \cite{Guo2020} and \cite{Xu2020}, we consider the single hop communication among APs. The signal to interference plus noise ratio (SINR) between AP ${m_1}$ and AP ${m_2}$ is calculated by
\begin{equation}
    \varphi_{m_1,m_2} = \frac{p_m^{a}|h_{m_1,m_2}|^2}{d_{m_1,m_2}\times \left (\sum_{m \neq m_1,m_2}^{\mathcal{M}}{\frac{p_m^{i}|h_{m,m_2}|^2}{d_{m,m_2}} } +\sigma^2\right )}.
\end{equation}

\par In the probability-based model, to calculate the transmission overhead of a \emph{Follower}, we adopt the mean of SINR between \emph{Follower} and all other consensus nodes as the SINR when a specific \emph{Follower} sends signals to the \emph{Leader}
\begin{equation}
    \overline{\varphi} = \frac{1}{M-1} \sum_{i \in \mathcal{M}, i \neq m}{\varphi_{m,i}}.
\end{equation}

\par It is worth noting that the data size of status confirmation is much smaller than that of the block\cite{Guo2020}. In order to make efficient use of limited bandwidth as much as possible, we assume that APs transmit two different data with fixed bandwidth according to the data size. Thus the bandwidth allocation strategy can be formulated as
\begin{subequations}\label{P}
	\begin{align}
		&\quad \frac{L^s}{B^{s}_m} = \frac{L^b}{B^{b}_m},\\
		s.t.\;
		&\quad B^{s}_m + B^{b}_m = \overline B_m
	\end{align}
\end{subequations}
where $B^{s}_m$, $B^{b}_m$ are the fixed bandwidth used by AP $m$ to transmit status confirmation signal and blocks, respectively. When the \emph{Leader} receives the block confirmation signals returned by the \emph{Followers}, the block will be admitted and attached to the blockchain. All the transaction information in this block will be completely transparent and cannot be changed.

\section{Problem Formulation and Analysis}

\subsection{Energy Consumption Analysis}

\subsubsection{Energy Consumption of Task Offloading}
 \par Since users will divide their tasks in proportion and transmit them to the corresponding AP in the cluster for processing, for user $n$, the data transmission delay when offloading computing task to AP $m$ can be given as
\begin{equation}
    D_{m,n}^{u} = \frac{a_{m,n}L_n}{r_{m,n}}.
\end{equation}

\par When AP receives the task, it will consume its own computing resources for task processing. The data processing delay when user $n$ offloading its task to  AP $m$ can be given as
\begin{equation}
    D_{m,n}^{e} = \frac{a_{m,n} L_n \rho_n}{c_{m,n}}.
\end{equation}

\par According to \cite{Zhao2017} and \cite{Wu2021}, the energy consumption of a device is positively correlated with the running time. When user $n$ transmit data to AP $m$, the energy consumption for uplink transmission can be formulated as
\begin{equation}
    E^{u}_{m,n} = \epsilon_{c} D^{u}_{m,n},
\end{equation}
where $\epsilon_{c}$ is the transmission energy consumption coefficient (in J/ms).

\par Similarly, the energy consumption of AP $m$ when processing task of user $n$ can be formulated as
\begin{equation}
    E_{m,n}^{e} =\epsilon_{p} D^{e}_{m,n},
\end{equation}
where $\epsilon_{p}$ represents the computing energy consumption coefficient.

\par Thus, the total offloading energy consumption of AP $m$ and user $n$  are expressed as
\begin{equation}
    E_m^{o} = \sum_{n \in \mathcal{N}} { E_{m,n}^{e} },
\end{equation}
\begin{equation}
    E_n^{o} = \sum_{m \in \mathcal{M}}{E_{m,n}^{u}}.
\end{equation}
 respectively.

\subsubsection{Energy Consumption of Consensus Mechanism}
\par In the election phase of R-RAFT, the \emph{Leader} needs to perform $(2+D^i)\left(M-1\right)$ times state confirmation communication, including $(M-1)$ times voting requests,  $\left(M-1\right)$ times election confirmations and $D^i\left(M-1\right)$ times heartbeats. In addition, the \emph{Leader} needs to perform  $\left (M-1\right)$ times block duplication in the block verification phase.
\par {\color{black}It's worth noting that APs need to occupy bandwidth and consume energy in each data transmission. Hence, the total energy consumption for data transmission should be equal to the sum of the energy consumed by each transmission. In addition, since APs are densely deployed and transmit data through wireless links in UC-MEC, the transmission rate of each AP can still be obtained through the Shannon formula. According to the bandwidth allocation strategy mentioned in part 2.2.1, the bandwidth of each state confirmation signal transmission is equal to ${B^{s}_{l}}/(2+D^i)$. Thus the energy consumption for the \emph{Leader} to transmit state confirmation signal can be expressed as}

        {\setlength{\parindent}{1em}
        \begin{equation}
            E_{l}^{s}= \sum_{i \in \mathcal{M},i\neq Leader} \epsilon_c \frac{(2+D^i)\left(M-1\right)\times L^s}{\frac{B^{s}_{l}}{2+D^i}\times log_2\left(1 + \varphi_{l,i}\right)}.
        \end{equation}
        }

\par {\color{black} Similarity, the energy consumption for the \emph{Leader} to duplicate block can be expressed as}

\begin{equation}
    E_{l}^{b}= \sum_{i \in \mathcal{M}, i\neq Leader} \epsilon_c \frac{\left(M-1\right)\times L^b}{\frac{B^{b}_{l}}{M-1}\times log_2\left( 1 + \varphi_{l,i}\right )}.
\end{equation}
\par Then the energy consumption of the \emph{Leader} for data transmission is formulated as
\begin{equation}
    E_{l}^{t} = E_{l}^{s} + E_{l}^{b},
\end{equation}

\par In addition, the \emph{Leader} needs to consume the computing resources to generate blocks. The computing delay required by the \emph{Leader} to perform block generation is represented by
\begin{equation}
    D_{l}^{g} = \frac{L^b}{\overline{C}_{l}},
\end{equation}

\par Then, energy consumption for \emph{Leader} to generate the block is
\begin{equation}
   E_{l}^{g} = \epsilon_p D_{l}^{g},
\end{equation}

\par During the consensus, each \emph{Follower} needs send a vote reply, $D^i$ heartbeat packet replies, and a block confirmation reply, thus total $\left(D^i+2\right)$ times state confirmation transmission. We can get the transmission delay of a \emph{Follower} as
\begin{equation}
    D_{f}^{s} =\frac{L^s\left(D^i+2\right)}{\frac{\overline{B}_m}{D^i+2} log_2 \left(1+\overline{\varphi}\right )  }  .
\end{equation}

\par Thus, the energy consumption of a \emph{Follower} for data transmission is
\begin{equation}
    E_{f}^{t} = \epsilon_c D_{f}^{s}.
\end{equation}
\par Therefore, the energy consumption of AP $m$ in the consensus is expressed as
\begin{equation}
  \begin{split}
     E_{m}^{c} & = \mathbb{E}\left [ E_{l} \right] + \mathbb{E}\left[E_{f} \right] \\
       & =  \left(E_{l}^{g}+E_{l}^{t}\right)\times P\left(m = Leader\right) \\
       & + \left(E_{f}^{t}\right)\times P\left(m=Follower\right).
  \end{split}
\end{equation}

\subsubsection{Total Energy Consumption}
\par In our system, total energy consumption includes offloading energy consumption of all users and offloading and consensus energy consumption of all APs. Therefore, the energy consumption of UC-MEC can be formulated as
\begin{equation}
  E^{a} = \sum_{m\in \mathcal{M}}\left(E_{m}^{o} + E_m^{c} \right)+ \sum_{n\in \mathcal{N}} E_{n}^{o}.
\end{equation}

\subsection{Problem Formulation}
During the offloading process, for user $n,n\in \mathcal{N}$, its tasks will be processed in parallel by all APs in the AP cluster. Hence, offloading delay of user $n$ is the maximum value of the service delay of all APs in the AP cluster, which is given by
\begin{equation}
    D_n^{o} = \max\{D_{1,n}^{o}, D_{2,n}^{o},...,D_{m,n}^{o}\}, \forall m \in \mathcal{M},
\end{equation}
where $D_{m,n}^{o}=D_{m,n}^{u}+D_{m,n}^{e}$.

\par We assume the clustering profiles of all APs is $\boldsymbol A =\left\{ a_{m,n}, \forall m \in \mathcal{M}, n \in \mathcal{N}  \right\}$. The computing resource allocation and bandwidth allocation profiles of all APs are denoted as $\boldsymbol C = \left\{ c_{m,n}, \forall m \in \mathcal{M}, n \in \mathcal{N}  \right\}$ and $\boldsymbol B = \left\{ b_{m,n}, \forall m \in \mathcal{M}, n \in \mathcal{N}  \right\}$, respectively. Formally, the energy consumption minimization problem in offloading and consensus is formulated as follows:
\begin{subequations}\label{P}
	\begin{align}
		\mathrm{\mathcal{P}1}:
		&\quad \min_{\boldsymbol A,\boldsymbol C,\boldsymbol B} E^{a},\\
		s.t.\;
		&\quad D_n^{o} \leq D_n^{t}, \forall n \in \mathcal{N}, \\
		&\quad \sum_{n=1}^{N}{c_{m,n}} \leq C_m, \forall m \in \mathcal{M}, \\
		&\quad \sum_{n=1}^{N}{b_{m,n}} \leq B_m, \forall m \in \mathcal{M}, \\
		&\quad a_{m,n}\in [0,1],\forall n\in \mathcal N,\forall m\in \mathcal M, \\
		&\quad \sum_{m=1}^{M}{a_{m,n}}=1, \forall n\in \mathcal N.
	\end{align}
\end{subequations}
where (26b) represents that the offloading delay cannot exceed the task delay threshold; (26c) and (26d) represent that the computing resources and the bandwidth allocated by an AP cannot exceed its computing capability and bandwidth; (26e) represents that the clustering variable is a continuous variable in $\left[0,1\right]$; (26f) represents that the task offloaded by each user will be fully processed by APs.

\section{Joint Optimization for APs clustering and Resource Allocation }

\subsection{ADMM-based Parallel Optimization Framework}
\subsubsection{Problem Transformation}
\par Since optimization variables in $\mathcal{P}1$ are coupled with each other, making $\mathcal{P}1$ a non-convex problem, it is necessary to introduce new auxiliary variables and constraints to decouple and simplify the problem. Expanding Eq. (24) and substituting it into $\mathcal{P}1$, the optimization problem can be reformulated as
\begin{subequations}
\begin{align}
    \min_{\boldsymbol A,\boldsymbol C,\boldsymbol B} & \sum_{m}^{\mathcal{M}} \bigg (\frac{1}{M}\sum_{n}^{\mathcal{N}}{\epsilon_c D_{m,n}^{u}} + \frac{1}{M} \sum_{n}^{\mathcal{N}}{\epsilon_p D_{m,n}^{e}} +  \left( 1-\frac{R_m}{R}\right)\times  \\
    &\epsilon_c D_{f}^{s} + \frac{R_m}{R} \big (\epsilon_p D_{l}^{g} + E_l^s+ E_l^b \big ) \bigg ) \nonumber \\
    & s.t. \left(26b\right)-\left(26f\right).
\end{align}
\end{subequations}
\par Observing the above optimization problem carefully, we can find the following coupling items:
\begin{itemize}
  \item $D_{m,n}^{u}$ and $D_{m,n}^{e}$ contain $\frac{a_{m,n}}{b_{m,n}}$ and $\frac{a_ {m,n}}{c_{m,n}}$, respectively;
  \item $\frac{\overline C_m + \overline B_m}{\overline B_m}$ which is equal to $\frac{C_m-\sum_n^{\mathcal{N}}c_{m,n}}{B_m-\sum_n^{\mathcal{N}}b_{m,n}}$ is included in $\frac{ R_m}{R}\left(E_{l}^{s} + E_{l}^{b}\right)$ and $\frac{\epsilon_c R_m}{R}D_{f}^{s}$;
  \item $\frac{\overline C_m + \overline B_m}{\overline C_m}$ is contained in $\frac{\epsilon_p R_m}{R} D_{l}^{g}$;
  \item There are coupling items $\frac{a_{m,n}}{c_{m,n}}$ and $\frac{a_{m,n}}{b_{m,n}}$ in constraint $\left(26b\right)$.
\end{itemize}
\par Let $b^{'}_{m,n}=\frac{1}{b_{m,n}}$ and $c^{'}_{m,n}=\frac{1}{c_{m,n}}$, $\forall m \in \mathcal{M},\forall n \in \mathcal{N}$, we can introduce the following auxiliary variables:
\begin{equation}
    \chi_{m,n} = a_{m,n}b^{'}_{m,n}, \forall m \in \mathcal{M}, \forall n \in \mathcal{N},
\end{equation}
\begin{equation}
    \psi_{m,n} = a_{m,n}c^{'}_{m,n},\forall m \in \mathcal{M}, \forall n \in \mathcal{N},
\end{equation}
\begin{equation}
    \kappa_{m,n} = \frac{C_m-\sum_n^{\mathcal{N}}\frac{1}{c^{'}_{m,n}}}{B_m-\sum_n^{\mathcal{N}}\frac{1}{b^{'}_{m,n}}},\forall m \in \mathcal{M}.
\end{equation}
\par According to reformulation linearization technology (RLT)\cite{Hou2010}, the above three equations can be replaced by the following three constraint sets:
\begin{equation}
\left\{
\begin{aligned}
&\chi_{m,n} \geq \frac{a_{m,n}}{B_m}, &\forall m \in \mathcal{M}, \forall n \in \mathcal{N}; \\
&\chi_{m,n} \leq b^{'}_{m,n}+\frac{a_{m,n}}{B_m}-\frac{1}{B_m}, &\forall m \in \mathcal{M}, \forall n \in \mathcal{N}; \\
&\chi_{m,n} \leq \frac{a_{m,n}}{\beta}, &\forall m \in \mathcal{M}, \forall n \in \mathcal{N}; \\
&\chi_{m,n} \geq \frac{a_{m,n}}{\beta}-\frac{1}{\beta}+b^{'}_{m,n}, & \forall m \in \mathcal{M}, \forall n \in \mathcal{N}.
\end{aligned}
\right.
\end{equation}
\begin{equation}
\left\{
\begin{aligned}
&\psi_{m,n} \geq \frac{a_{m,n}}{C_m}, &\forall m \in \mathcal{M}, \forall n \in \mathcal{N}; \\
&\psi_{m,n} \leq c^{'}_{m,n}+\frac{a_{m,n}}{C_m}-\frac{1}{C_m}, &\forall m \in \mathcal{M}, \forall n \in \mathcal{N}; \\
&\psi_{m,n} \leq \frac{a_{m,n}}{\beta}, &\forall m \in \mathcal{M}, \forall n \in \mathcal{N}; \\
&\psi_{m,n} \geq \frac{a_{m,n}}{\beta}-\frac{1}{\beta}+c^{'}_{m,n}, & \forall m \in \mathcal{M}, \forall n \in \mathcal{N}.
\end{aligned}
\right.
\end{equation}
\begin{equation}
\left\{
\begin{aligned}
&\kappa_{m} \geq \frac{C_m-\sum_n^{\mathcal{N}}\frac{1}{c^{'}_{m,n}}}{B_m},&\forall m \in \mathcal{M};\\
&\kappa_{m} \leq \frac{C_m}{B_m-\sum_n^{\mathcal{N}}\frac{1}{b^{'}_{m,n}}}-\frac{C_m}{B_m},&\forall m \in \mathcal{M}; \\
&\kappa_{m} \leq \frac{C_m-\sum_n^{\mathcal{N}}\frac{1}{c^{'}_{m,n}}}{\beta},&\forall m \in \mathcal{M};\\
&\kappa_{m} \geq \frac{C_m}{B_m-\sum_n^{\mathcal{N}}\frac{1}{b^{'}_{m,n}}}-\frac{\sum_n^{\mathcal{N}}\frac{1}{c^{'}_{m,n}}}{\beta},&\forall m \in \mathcal{M}.
\end{aligned}
\right.
\end{equation}
where $\beta$ is an infinitesimal positive number.
\par  It's easy to prove that the above three constraint sets are equivalent to equations (28) - (30) according to\cite{Zhao2017} and \cite{Wang2019}. In addition, constraint (26b) can also be transformed into the following form:

\begin{equation}
    \frac{L_n \chi_{m,n}}{log_2(1+r_{m,n})} + L_n \rho_n \psi_{m,n} \leq D_n^{t}, \forall m \in \mathcal{M},\forall n \in \mathcal{N}
\end{equation}
\par Constraints (26c), (26d) can be transformed to the following forms:
\begin{equation}
    \sum_n^{\mathcal{N}}\frac{1}{c^{'}_{m,n}} \leq C_m, \forall m \in \mathcal{M}
\end{equation}
\begin{equation}
    \sum_n^{\mathcal{N}}\frac{1}{b^{'}_{m,n}} \leq B_m, \forall m \in \mathcal{M}
\end{equation}

\par Substituting (28) - (36) into (27), the optimization objective can be expressed as the sum of $M$ functions in the same form. Therefore, the optimization problem $\mathcal{P}1$ can be reformulated as
\begin{subequations}\label{P2}
	\begin{align}
		\mathrm{\mathcal{P}2}:
		&\quad \min_{\boldsymbol A,\boldsymbol C^{'},\boldsymbol B^{'},\boldsymbol{X},\boldsymbol{\Psi},\boldsymbol{K}} \sum_{m}^{\mathcal{M}}V_m(\boldsymbol{B^{'}},\boldsymbol{X},\boldsymbol{\Psi},\boldsymbol{K})\\
		s.t.\;
		&\quad (26e),(26f),(31)-(36).
	\end{align}
\end{subequations}

\par As far as \emph{P2} is concerned, the numbers of variables and constraints reach $3MN + M$ and $9MN + 6M$, respectively. When the number of users and APs increases, the worse-case complexity will increase exponentially, and the time to obtain the optimal solution will be unbearable. As a mature optimization framework, ADMM \cite{Boyd} can be the reliable solution for solving large-scale convex or non-convex problems.
{\color{black}Through the transformation of ADMM method, the originally complex problem can be decomposed into the sum of several simpler sub-problems, each sub-problem can be allocated to a node for parallel processing. Compared with other classical centralized algorithms, ADMM can significantly reduce the computational complexity and running time. In this paper, we develop a novel ADMM-based parallel optimization framework to solve \emph{P2}. To reduce complexity, The proposed framework decompose the original problem into $M$ same sub-problems and each AP can solve a sub-problem in parallel.}

\subsubsection{The Update of Global Variables}
Clearly, the existence of constraint (26f) makes each AP need to know the global clustering information (clustering decision of other APs), so that the problem cannot be decomposed. Thus, we introduce a local copy of $a$, denoted as $\hat{a}$, which satisfies the following formula
\begin{equation}
    \hat a_{m,n} = a_{m,n}, \forall m \in \mathcal{M}, \forall n \in \mathcal{N}.
\end{equation}
\par Then the constraint (26f) is transformed into the following form
\begin{equation}
    \sum_m^{\mathcal{M}} \hat a_{m,n} = 1, \forall n \in \mathcal{N}.
\end{equation}
\par Thus, the optimization problem can be further reformulated as
\begin{subequations}\label{P3}
	\begin{align}
		\mathrm{\mathcal{P}3}:
		&\quad \min_{\hat {\boldsymbol A},\boldsymbol C^{'},\boldsymbol B^{'},\boldsymbol{X},\boldsymbol{\Psi},\boldsymbol{K}} \sum_{m}^{\mathcal{M}}V_m(\boldsymbol{A},\boldsymbol{B^{'}},\boldsymbol{X},\boldsymbol{\Psi},\boldsymbol{K})\\
		s.t.\;
		&\quad (26e),(31)-(36),(38),(39).
	\end{align}
\end{subequations}
\par Let $\boldsymbol \Pi = \{\boldsymbol A,\boldsymbol C^{'},\boldsymbol B^{'},\boldsymbol{X},\boldsymbol{\Psi},\boldsymbol{K}\}$ and $\ddot{\boldsymbol{\Pi}}=\{\boldsymbol{A},\boldsymbol B^{'},\boldsymbol{X},\boldsymbol{\Psi},\boldsymbol{K} \}$. Then, the augmented Lagrangian function of $\mathcal{P}3$ is represented as
\begin{equation}
\begin{aligned}
 \mathcal{L}(\ddot{\boldsymbol {\Pi}},\hat{\boldsymbol{A}},\boldsymbol \lambda) & = \sum_{m}^{\mathcal{M}}V_m(\ddot{\boldsymbol {\Pi}}) + \sum_{m}^{\mathcal{M}} \sum_{n}^{\mathcal{N}} \lambda_{m,n} (a_{m,n} - \hat a_{m,n}) \\
  & + \frac{q}{2}\sum_{m}^{\mathcal{M}} \sum_{n}^{\mathcal{N}} (a_{m,n} - \hat a_{m,n})^2
\end{aligned}
\end{equation}
where $\boldsymbol{\lambda} = \left\{\lambda_{m,n}, \forall m \in \mathcal{M}, \forall n \in \mathcal{N}\right\}$ is the Lagrange multiplier, and $q\geq 0$ is the augmented Lagrangian parameter, which affects the convergence of ADMM algorithm.

\par According to \cite{Liu2018}, \cite{Park}, the global variable $\boldsymbol \Pi$ in $\mathcal{P}3$ is updated at the $\left(t+1\right)$-th iteration by solving the following optimization problem:
\begin{subequations}\label{P4}
	\begin{align}
		\mathrm{\mathcal{P}4}:
		&\quad \min_{\boldsymbol \Pi} \mathcal{L}(\ddot{\boldsymbol{\Pi}},\hat {\boldsymbol A}^{(t)}, \boldsymbol\lambda^{(t)})\\
		s.t.\;
		&\quad (26e),(31)-(36).
	\end{align}
\end{subequations}
where $\left(t\right)$ represents the index of the number of iterations.

\par {\color{black} For each AP $m$, we define $\boldsymbol a_m =\{ a_{m,n},\forall n \in \mathcal{N} \}$ as the clustering decision vector, $\boldsymbol {b^{'}_m} =\{ b^{'}_{m,n},\forall n \in \mathcal{N} \}$ as the bandwidth allocation vector and $\boldsymbol {c^{'}_m} =\{ c^{'}_{m,n},\forall n \in \mathcal{N} \} $ as the computing resource allocation vector, respectively. Similarly, we define $\boldsymbol \chi_m =\{ \chi_{m,n},\forall n \in \mathcal{N} \}$, $\boldsymbol \psi_m =\{ \psi_{m,n},\forall n \in \mathcal{N} \}$, $\boldsymbol{\pi_m} = \{ \boldsymbol a_m, \boldsymbol b_m^{'},\boldsymbol c_m^{'}, \boldsymbol \chi_m,\boldsymbol \psi_m,\boldsymbol \kappa_m  \}$, and  $\ddot{\boldsymbol{\pi}}_{m}=\{  \boldsymbol a_m,\boldsymbol b^{'}_m,\boldsymbol \chi_m,\boldsymbol \psi_m,\boldsymbol \kappa_m\}$. Then, we can rewrite the augmented Lagrangian function of $\mathcal{P}4$ as}
\begin{equation}
    \mathcal{L}(\ddot{\boldsymbol{\Pi}},\hat {\boldsymbol A}^{(t)}, \boldsymbol \lambda^{(t)}) = \sum_{m}^{\mathcal{M}}\mathcal {F}_m (\ddot{\boldsymbol{\pi_m}}).
\end{equation}
\par Secondly, by analyzing the problem $\mathcal{P}4$, we can get that the objective function of $\mathcal{P}4$ is convex for all variables. However, the second and the fourth item of constraint (33) are non-convex, making the problem still intractable. Therefore, we take a relaxation for (33) by adding it into the objective function of $\mathcal{P}4$ in the form of a penalty term, which can be expressed as
\begin{equation}
    \begin{aligned}
          \Upsilon(\boldsymbol K,\boldsymbol B^{'},\boldsymbol C^{'}) & = \zeta_1\left(\frac{C_m-\sum_{n}^{\mathcal{N}}\frac{1}{c^{'}_{m,n}}}{B_m}-\kappa_m\right) \\
          & + \zeta_2\left(\kappa_m-\frac{C_m}{B_m-\sum_n^{\mathcal{N}} \frac{1}{b_{m,n}^{'}} } + \frac{C_m}{B_m}\right).
    \end{aligned}
\end{equation}
where $\zeta_1$ and $\zeta_2$ are penalty coefficients. It's easy to prove that when the values of $\zeta_1$ and $\zeta_2$ are large enough, the global optimization problem $\mathcal{P}4$ can be transformed into the following form
\begin{subequations}\label{P4}
	\begin{align}
		\mathrm{\mathcal{P}4^{'}}:
		&\quad \min_{\boldsymbol \Pi} \sum_{m}^{\mathcal{M}}\mathcal {F}_m \left(\ddot{\boldsymbol{\pi}}_{m}\right) - \Tilde{\Upsilon}\left(\boldsymbol K,\boldsymbol B^{'},\boldsymbol C^{'}\right)\\
		s.t.\;
		&\quad (26e),(31)-(36),
	\end{align}
\end{subequations}
where $\Tilde {\Upsilon}(\boldsymbol K,\boldsymbol B^{'},\boldsymbol C^{'}) = -\Upsilon(\boldsymbol K,\boldsymbol B^{'},\boldsymbol C^{'})$. Obviously, $\Tilde {\Upsilon}(\boldsymbol K,\boldsymbol B^{'},\boldsymbol C^{'})$ is a convex function. To sum up, the optimization problem of updating the global variable $\boldsymbol{\pi_m}$ independently and in parallel for each AP $m$ is expressed as
\begin{subequations}\label{P5}
	\begin{align}
		\mathrm{\mathcal{P}5}:
		&\quad \min_{\boldsymbol \pi_m} \mathcal {F}_m (\ddot{\boldsymbol{\pi}}_{m}) - \Tilde{\Upsilon_m}(\boldsymbol \kappa_m,\boldsymbol b^{'}_{m},\boldsymbol c^{'}_{m})\\
		s.t.\;
		&\quad (26e),(31)-(36).
	\end{align}
\end{subequations}
\par The optimization objective of $\mathcal{P}5$ is the difference of two convex functions, which is a D.C. Programming Problem\cite{Vucic}. Similar to \cite{Wang2019}, we propose a sub-gradient iteration algorithm to solve $\mathcal{P}5$. The details of the proposed algorithm are summarized in \textbf{Algorithm 1}.

\par In \textbf{Algorithm 1}, $\nabla$ denotes the sub-gradient of the corresponding variable of the objective function, and $\Tilde{\Upsilon}^{(n)} = \Tilde{\Upsilon}(\boldsymbol \kappa_m^{(n)},\boldsymbol b^{'(n)}_{m},\boldsymbol c^{'(n)}_{m})$.Since $\mathcal{P}5^{'}$ is convex, and all constraints of it are linear, we can use convex optimization tools such as CVX\cite{cvx} to solve it.

\subsubsection{The Update of Local Variables}
\par After each AP independently solves the global variables in parallel, the results will be aggregated to a central node. According to \cite{Chen2021}, we can set up an anchor AP, and the integrated MEC  on the anchor AP can be the central node. After the central node collects and integrates the global variables, it will be responsible for updating the local variables in the $\left(t+1\right)$-th iteration. The updating of local variables $\hat{\boldsymbol A}$ depends on solving the following optimization problem
\begin{subequations}\label{P6}
	\begin{align}
		\mathrm{\mathcal{P}6}:
		&\quad \min_{\hat{\boldsymbol A}} \mathcal{L}(\hat {\boldsymbol A},\hat{\boldsymbol{\Pi}}^{(t+1)},\boldsymbol \lambda^{(t)})\\
		s.t.\;
		&\quad (39),\hat a_{m,n} \in [0,1],\forall m \in \mathcal{M},\forall n \in \mathcal{N}.
	\end{align}
\end{subequations}
\par After removing some constants, $\mathcal{P}6$ is equivalent to solving the following problem
\begin{subequations}\label{$P6^{'}$}
	\begin{align}
		\mathrm{\mathcal{P}6^{'}}:
		&\quad \min_{\hat{\boldsymbol A}} \hat{\mathcal{J}}(\hat{\boldsymbol A})\\
		s.t.\;
		&\quad (39),\hat a_{m,n} \in [0,1],\forall m \in \mathcal{M},\forall n \in \mathcal{N}.
	\end{align}
\end{subequations}

where
\begin{equation}
    \begin{aligned}
        \hat{\mathcal{J}}(\hat{\boldsymbol A}) & = \sum_m^{\mathcal{M}}\sum_{n}^{\mathcal{N}} \lambda_{m,n}^{(t)} \left(a_{m,n}^{(t+1)} - \hat a_{m,n}\right) \\
       & + \frac{q}{2} \sum_m^{\mathcal{M}}\sum_{n}^{\mathcal{N}} \left(a_{m,n}^{(t+1)} - \hat a_{m,n}\right)^2.
    \end{aligned}
\end{equation}
\par Obviously, the problem $\mathcal{P}6^{'}$ is also convex. Therefore, the local variables $\hat{\boldsymbol A}$ can be solved with some mature convex optimization tools.

\begin{algorithm}[tbp]
	\renewcommand{\algorithmicrequire}{\textbf{Input:}}
	\renewcommand{\algorithmicensure}{\textbf{Output:}}
	\caption{Sub-gradient Based Iteration Algorithm}
	\label{alg:1}
	\begin{algorithmic}[1]
		\REQUIRE Initial a feasible value $\left\{\boldsymbol k_m^{(n)}, \boldsymbol b_{m}^{'(n)},\boldsymbol c_{m}^{'(n)} \right\}$for AP $m$ where $m\in\mathcal{M}$ and $n=0$.
		\ENSURE The global variable $\boldsymbol \pi_m$ for AP $m$ where $m \in \mathcal{M}$.
        \REPEAT
		\STATE Solve the following convex problem named $\mathcal{P}5^{'}$:\\
\begin{subequations}\label{P5}
	\begin{align}
	\min_{\boldsymbol \pi_m}  &\bigg\{\mathcal {F}_m (\ddot{\boldsymbol{\pi}}_{m}) - \nabla_{b_{m}^{'}} \Tilde{\Upsilon}^{(n)}(b^{'}_{m}-b_{m}^{'(n)})\nonumber \\
        &-\nabla_{c^{'}_{m}} \Tilde{\Upsilon}^{(n)}(c^{'}_{m}-c_{m}^{'(n)}) - \nabla_{\kappa_m}\Tilde{\Upsilon}^{(n)}\nonumber \\
        &\times(\kappa_m-\kappa_m^{(n)}) - \Tilde{\Upsilon}^{(n)}\bigg\}\nonumber\\
		s.t.\;
		&\quad (26e),(31)-(36)\nonumber.
	\end{align}
\end{subequations}
        \STATE Obtain the optimal solution for the $\left(n+1\right)$-th iteration $\left\{\boldsymbol \pi_m \right\}^{(t+1)}$.
        \STATE $n = n+1$.
        \UNTIL{Global variable $\boldsymbol \pi_m$ converges}.
	\end{algorithmic}
\end{algorithm}

\subsubsection{The Update of Dual Variables}
\par After the global variables and local variables are obtained, the update of dual variables in the $\left(t+1\right)$-th iteration can be expressed as
\begin{equation}
    \lambda_{m,n}^{(t+1)} = \lambda_{m,n}^{(t)} + q\left(a_{m,n}^{(t+1)} - \hat a_{m,n}^{(t+1)}\right),\forall m \in \mathcal{M},\forall n \in \mathcal{N}.
\end{equation}
\par In a word, the efficient APs clustering and resource allocation strategy are obtained through the sequential iteration of global variables, local variables, and dual variables. The ADMM-based parallel optimization algorithm is summarized in \textbf{Algorithm 2}.

\subsection{Performance Analysis}
\subsubsection{Complexity Analysis}
\par In our proposed ADMM-based parallel iteration optimization algorithm, i.e. \textbf{Algorithm 2}, each AP is able to solve $\mathcal{P}5$ independently and in parallel, which greatly reduces the computational complexity. Moreover, for each AP, $\mathcal{P}5$ can be viewed as a D.C Programming Problem and solved by a sub-gradient iterative algorithm, as described in \textbf{Algorithm 1}. In $\mathcal{P}5^{'}$, there are $5N+1$ optimization variables and $10N+2$ convex constraints for each AP. We assume that $\mathcal{P}5^{'}$ requires $n^{d}$ iterations to converge. Then, the complexity of $\mathcal{P}5^{'}$ is $\mathcal{O}(\max_{m \in \mathcal{M}} (5N+1)(10N+2)n^{D}) = \mathcal{O}(\max_{m \in \mathcal{M}} N^2)$. Similarly, for the optimization problem $\mathcal{P}6$, the number of optimization variables is $MN$, and the number of constraints is $N + MN$. Assuming that $n^{l}$ steps are needed to solve the optimization problem, then its complexity is $\mathcal{O}(MN(N+MN)n^{l}) = \mathcal{O}(n^{l}(M^2N^2+MN^2))$. For the updating of dual variables, we assume that $n^{d}$ steps are required to solve (51), then its computational complexity is $\mathcal{O}(n^{d}MN)$. Assuming that \textbf{Algorithm 2} requires $t^{a}$ times to converge, then the complexity of overall algorithm is $\mathcal{O}(N^2+n^{l}(M^2N^2+MN^2)+n^{d}MN) = \mathcal{O}(M^2N^2)t^{a}$, which is greatly reduced compared using centralized algorithm to slove $\mathcal{P}1$ directly.

\begin{algorithm}[tbp]
	\caption{ADMM-Based Parallel Iteration Optimization Algorithm}
	\label{alg1}
	\begin{algorithmic}[h]
	    \STATE $\mathbf{Initialization:}$
	    \STATE 1. $t=1$, $\lambda^{(t)}=\mathbf{0}$ and $t_{max}=100$.
	    \STATE 2. Select a suitable iterative stopping conditions $\gamma$ of the MEC anchor. And choose a penalty factor $q$ of the ADMM algorithm and send it to all APs.
		\STATE 3. Each AP $m$ determines an initial feasible solution $\hat{\boldsymbol{A}}^{[1]}$ satisfying (26f)
		\WHILE {$|\boldsymbol A-\hat{\boldsymbol A}|_2 \geq \gamma$ and $t \leq t_{max}$}
		\STATE 1. Update global variables: each AP $m$ $\left(m\in \mathcal{M}\right)$ executes \textbf{Algorithm 1} to obtain the global variable $\boldsymbol{\Pi}^{(t+1)}$ and uploads it to the anchor node, respectively.
		\STATE 2. Update local variables: the anchor node updates the local variable $\hat {\boldsymbol{A}}^{(t+1)}$  based on (48).
		\STATE 3. Update dual variables: the dual multiplier $\lambda^{(t+1)}$ is updated based on (51).
		\STATE 4. Update $t=t+1$
		\ENDWHILE
		\STATE $\mathbf{Output:}$ the optimal clustering and resource allocation solution $\left\{ \boldsymbol \Pi  \right\}^{*}$.
	\end{algorithmic}
\end{algorithm}

\subsubsection{System delay Analysis}
\par {\color{black}  For delay-sensitive applications, the tasks offload to the edge server needs to be completed within a limited time, i.e., the offloading delay cannot be greater than the tolerance delay of the tasks. Furthermore, all APs need to complete the consensus process before the next round of offloading. Therefore, the system delay is also need to be considered in the blockchain-enabled UC-MEC system. As mentioned above, the offloading delay of user $n$ has given by Eq. (25). For the consensus delay, the data transmission between \emph{Leader} and \emph{Follower} are also in parallel, thus the consensus delay of the proposed system is}
\begin{equation}
    \begin{aligned}
    D^{c} & =\max_{m \in \mathcal{M}} \left(\max_{j \in \mathcal{M},j\neq m}\frac{\left(M-1\right) L^s\left(D^i+2\right)^2}{B_m^{s} \mathrm{log_2}\left(1+\varphi_{m,j}\right)}  \right.\\
    & \left. + \max_{j \in \mathcal{M},j\neq m}\frac{\left(D^i+2\right)^2 L^s}{\overline{B}_j \mathrm{log_2}\left(1+\overline{\varphi}\right)}  \right.\\
    & \left. +\max_{j \in \mathcal{M},j\neq m} \frac{\left(M-1\right)L^b}{B_m^{b}\mathrm{log_2}\left( 1 + \varphi_{m,j}  \right)} + D_m^{g} \right).
    \end{aligned}
\end{equation}

{\color{black} \par Therefore. the total delay of user $n$ is equal to the sum of the offloading delay and the consensus delay.}

\section{Simulation Results and Discussion}
In this section, the simulation results and discussion are presented to evaluated the performance of our algorithm. First, the simulation settings and comparison schemes are presented. Then it follows the convergence, delay, and energy consumption performance with various parameters. {\color{black} The simulation code is available on \href{https://github.com/qlt315/Blockchain-enabled-UCMEC}{https://github.com/qlt315/Blockchain-enabled-UCMEC}.  }

\subsection{Simulation Settings}

\par {\color{black}  We implement the simulations using MATLAB R2021b on a computer with a AMD Ryzen 3600 CPU running on a processor speed of 3.6 GHz, and 16 GB RAM. Assume that all APs and users are uniformly distributed in a $200$ m $\times$ $200$ m circle and the path loss model is $128.1+37.6$log$d$ ($d$ in kilometers) dB\cite{}. To better simulate the heterogeneity of APs and the diversity of offloading tasks, the bandwidth of APs are evenly distributed in [10,50] MHz\cite{Wang2019}, and the computing resources are evenly distributed in $[5,20]$ GHz \cite{Wu2021}. The size of user's task, the computing density, and the maximum tolerated delay are uniformly distributed in $[100, 200]$ kbits, $[1000,2000]$ CPU cycles/bit, and $[100,150]$ milliseconds, respectively. During the offloading and the consensus process, the transmission power of user, the transmission power of AP, and the interference signal power of other APs are $100$ mW, $200$ mW, $20$ mW \cite{Xu2022}, respectively. We assume the energy consumption coefficient of unit computing delay and transmission delay are $1$ J/s and $1.5$ J/s respectively, the power of Gaussian white noise is $-174$ dBm/Hz \cite{Liu2018}, and the size of state confirmation data is $10$ kbits. In the blockchain system, the block size is generally no more than 1 kbits. We assume that the block size is $500$ kbits and the block interval is $1$ second. In the proposed ADMM based algorithm, we assume the maximum tolerance error is $0.01$\cite{Wang2019} and the maximum number of iterations is $100$.}

\subsection{Comparison Schemes}
\par In order to better reflect the effectiveness of our proposed scheme, we also simulate the following three schemes for comparison. {\color{black} It should be noted that all the comparison schemes are deployed with blockchain to ensure the fairness of comparison.}
\begin{itemize}
  {\color{black}\item \textbf{Single Offloading (SO)}: To verify the effectiveness of the user-centric cooperative offloading method compared with traditional cellular-based MEC, each user will offload computing task to a single AP in this scheme[33]. Hence, the clustering decision variable $\mathcal{A}$ satisfies $\sum_{m}^{\mathcal{M}}a_{m,n}=1,\forall m \in \mathcal{M},\forall n \in \mathcal{N}$, $a_{m,n} \in \{0,1\}$.}
  
  {\color{black}\item \textbf{Block Coordinate Descent-based Offloading (BCDO)}: In this scheme, to verify the effectiveness of the proposed distributed ADMM optimization framework, we use centralized block coordinate descent (BCD) \cite{Chen2020,Bai2020}, to solve $\mathcal{P}2$ . Relying on
  the BCD technique, the original problem is decoupled into three sub-problems for alternatively optimizing the APs clustering, computing resource and bandwidth allocation decisions.}
  
  \item \textbf{Optimal Offloading (OO)}[35]: To verify the necessity of joint optimization of the offloading process and the consensus process, the optimization objective in this scheme only includes offloading energy consumption of all APs and users, i.e., $\sum_{m \in \mathcal{M}}E^{o}_m + \sum_{n \in \mathcal{N}}E^o_n$
  
  \item \textbf{RAFT-based Offloading (RO)}: To verify the effectiveness of our proposed R-RAFT consensus mechanism, we select RAFT consensus algorithm for comparison, thus the probability of each node becoming the \emph{Leader} is completely stochastic. We assume that $p_m=\emph{Leader}$ is a random number uniformly distributed in $\left(0,1\right]$.
\end{itemize}

\begin{figure*}[tbp]
    \centering
    \captionsetup{justification=centering}
	  \subfloat[Convergence performance under different AP numbers ($q=100$).]{
        \includegraphics[width=0.47\linewidth]{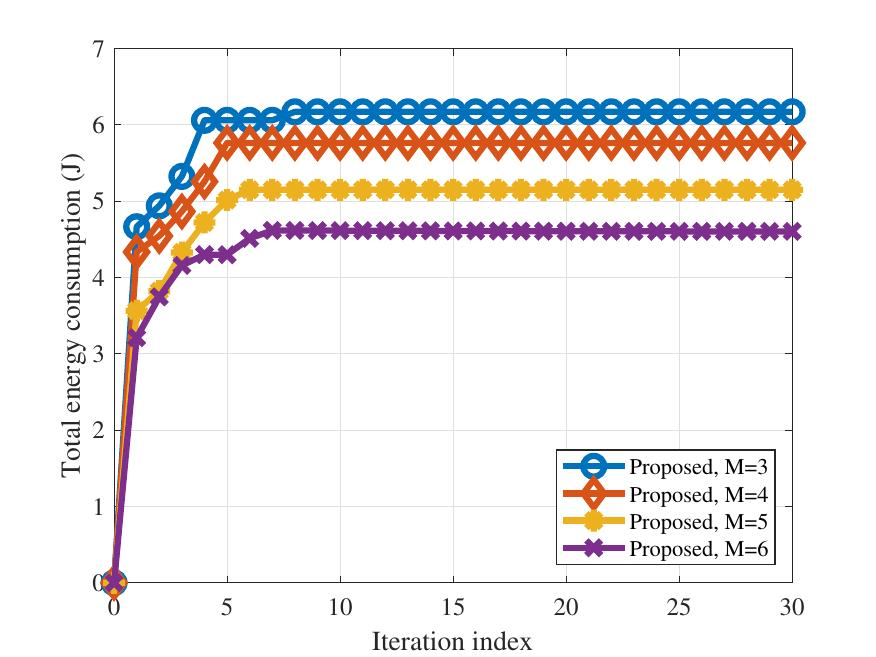}}
    \label{1a}
	  \subfloat[Convergence performance under different Lagrangian penalty factor $q$ ($M=5$).]{
        \includegraphics[width=0.47\linewidth]{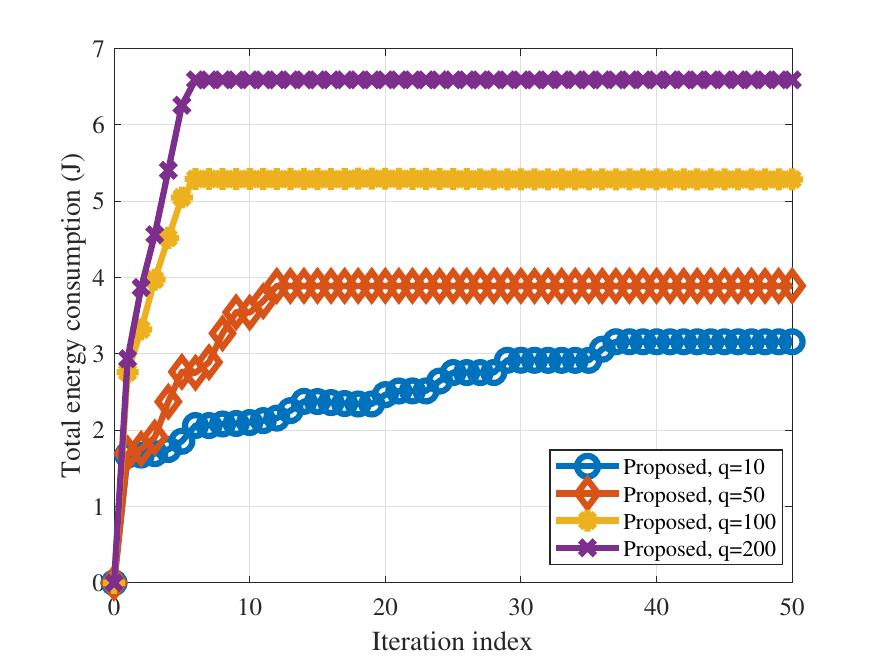}}
    \label{1b}
        \caption{Convergence performance comparison. }
	\end{figure*}

\subsection{Simulation Results}
\subsubsection{Convergence Performance}
\par We assume that the number of users $N=5$. The convergence processes of our proposed joint optimization scheme based on ADMM are shown in Fig. 3 (a) and Fig. 3 (b). We separately validate the effects of the Lagrangian penalty factor $q$ and the number of APs $M$ on the convergence performance. In addition, We compare the actual running times of BCD-based scheme and the propsoed ADMM-based scheme under different parameter settings in Fig. 3 (c).

\par As shown in Fig. 3 (a), the proposed scheme can rapidly converge within 10 steps. As the number of APs increases, the total energy consumption will decrease, but the downward trend will slow down. This is because the increase of APs  will enlarge the available computing and bandwidth resources in the network, thus reducing the computing and transmission delay and energy consumption. However, the number of communications among APs in consensus will also increase accordingly, so the consensus energy consumption will increase and gradually offset the gain of the reduction of the other parts of the energy consumption. Therefore, MEC service providers and communication service operators need to carefully decide the number or deploy density of AP.

\par It can be seen from Fig. 3 (b) that the value of the Lagrangian penalty factor $q$ will affect the convergence performance of the ADMM algorithm. When $q$ increases, the speed of convergence will increase, however, the total energy consumption will also increase. This is because the local copy of the clustering decision variable $\hat{\boldsymbol A}$ will approach the global variable at a faster speed when $q$ increases, thereby approximating the stopping condition of the algorithm faster. However, the increase in the convergence speed will sacrifice the quality of the solution, making the total energy consumption becomes relatively large. Therefore, the value of $q$ should be chosen carefully to trade off the convergence speed and the performance of the algorithm.

\begin{figure*}[tbp]
    \centering
    \captionsetup{justification=centering}
	  \subfloat[Average transmission rate under different user numbers.]{
        \includegraphics[width=0.47\linewidth]{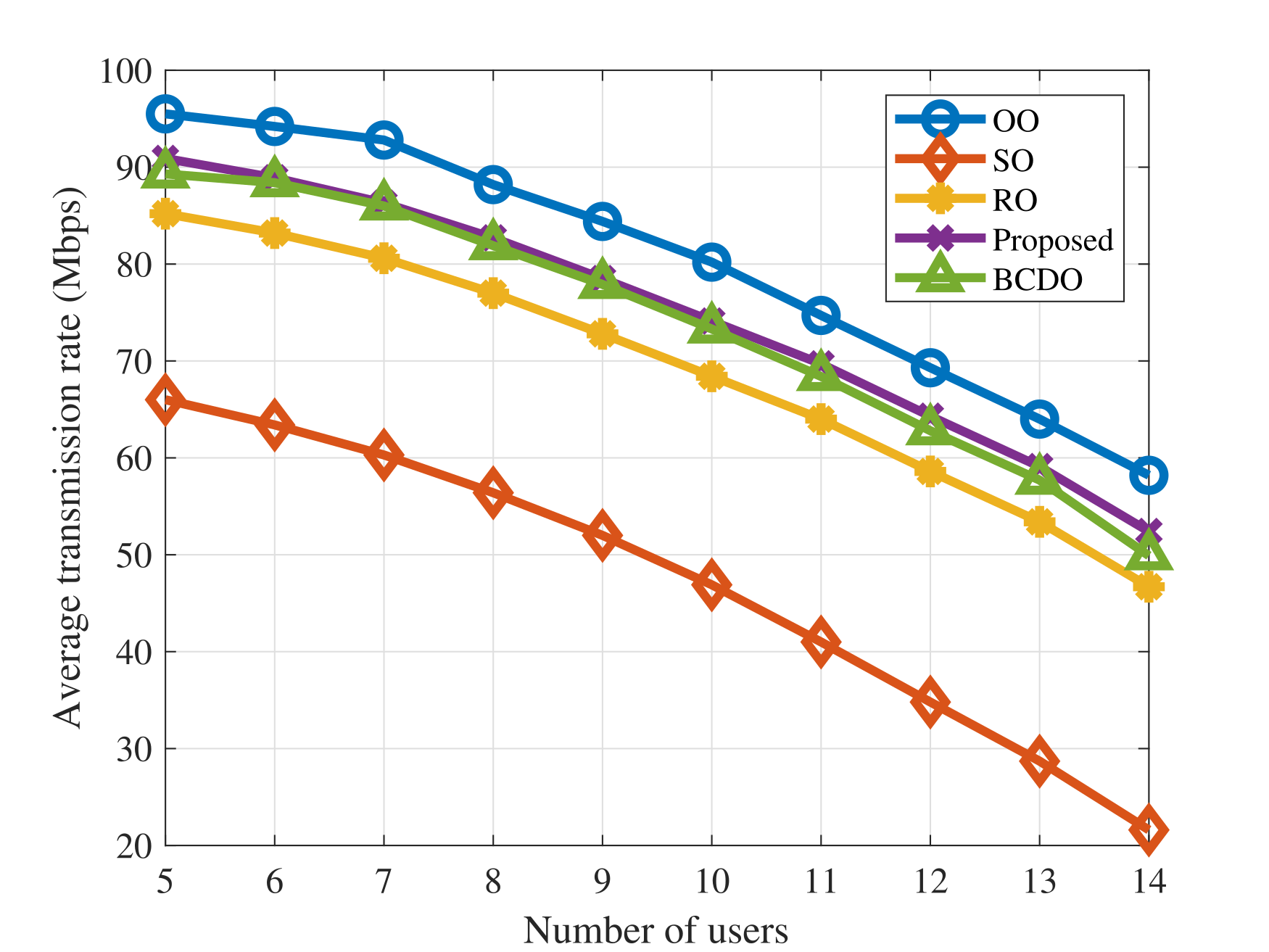}}
    \label{1a}
	  \subfloat[Average transmission rate under different AP numbers.]{
        \includegraphics[width=0.47\linewidth]{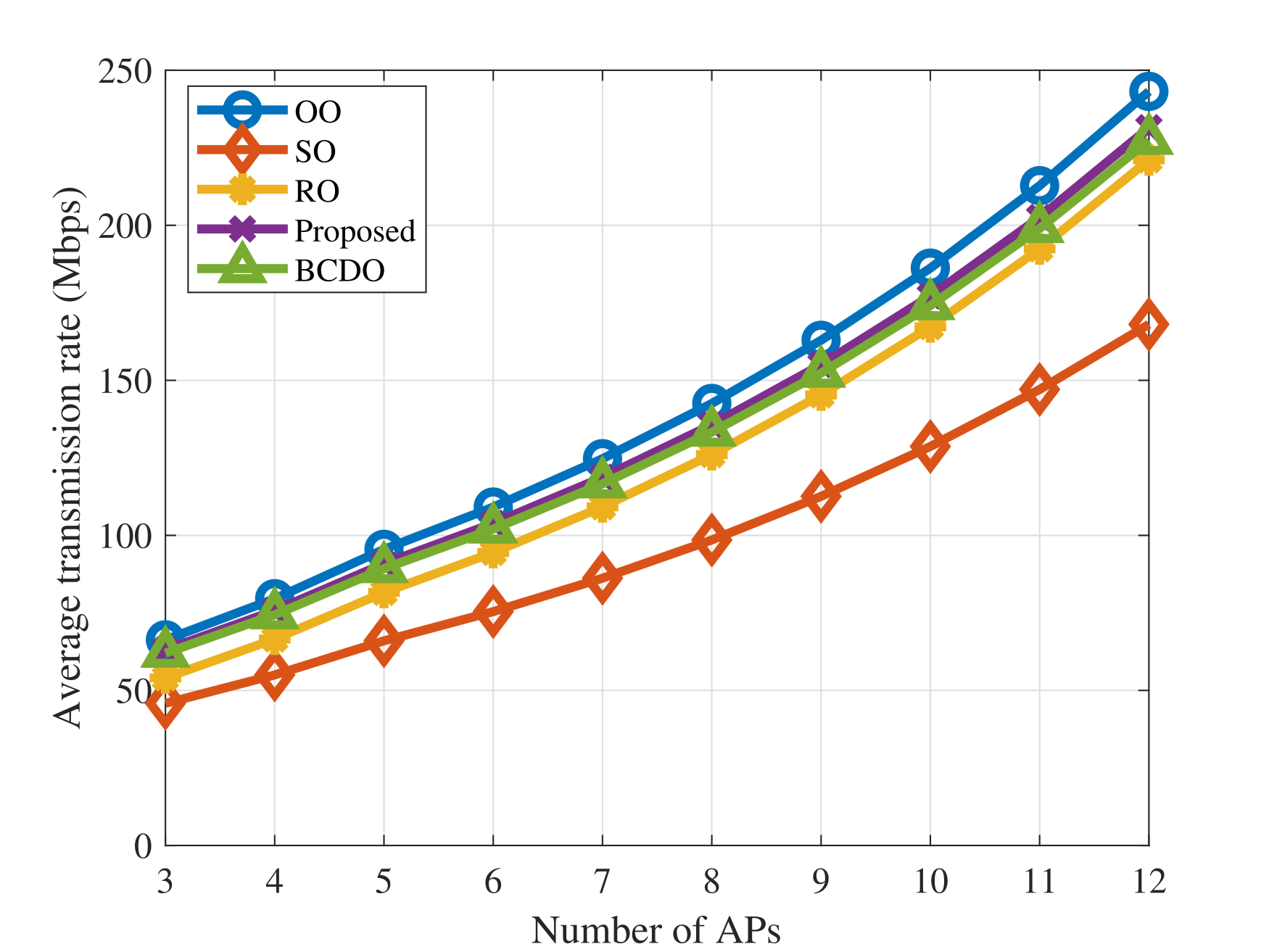}}
        \caption{Average transmission rate performance under different user and AP numbers.}
	\end{figure*}

\subsubsection{Uplink Transmission Rate Performance}
\par {\color{black}We now focus on the evaluation of the average uplink transmission rate of users. If not specifically mentioned, the number of users and APs are set to $5$, and the settings of other parameters are consistent with TABLE 2. Fig. 4 shows the average uplink transmission rate versus the number of users and APs. From Fig. 4 (a), we can see that when the number of users increases, the average transmission rate of all users will decrease. This is because the bandwidth resources allocated to each user will decrease and the inter-cluster interference will increase. As can be seen from Fig. 4 (b), when the number of APs increases, the available bandwidth resources of the system will increase, and the average transmission rate of users will be greatly improved. In addition, it can be seen that \textbf{SO} has a lower transmission rate than other schemes, which shows that the user-centric cooperative transmission mode is superior to the traditional transmission mode. The performance of our scheme is second only to \textbf{OO}, and \textbf{BCDO} has the same performance as the proposed scheme.}

\begin{figure*}[tbp]
    \centering
    \captionsetup{justification=centering}
	  \subfloat[Offloading delay under different user numbers.]{
        \includegraphics[width=0.3\linewidth]{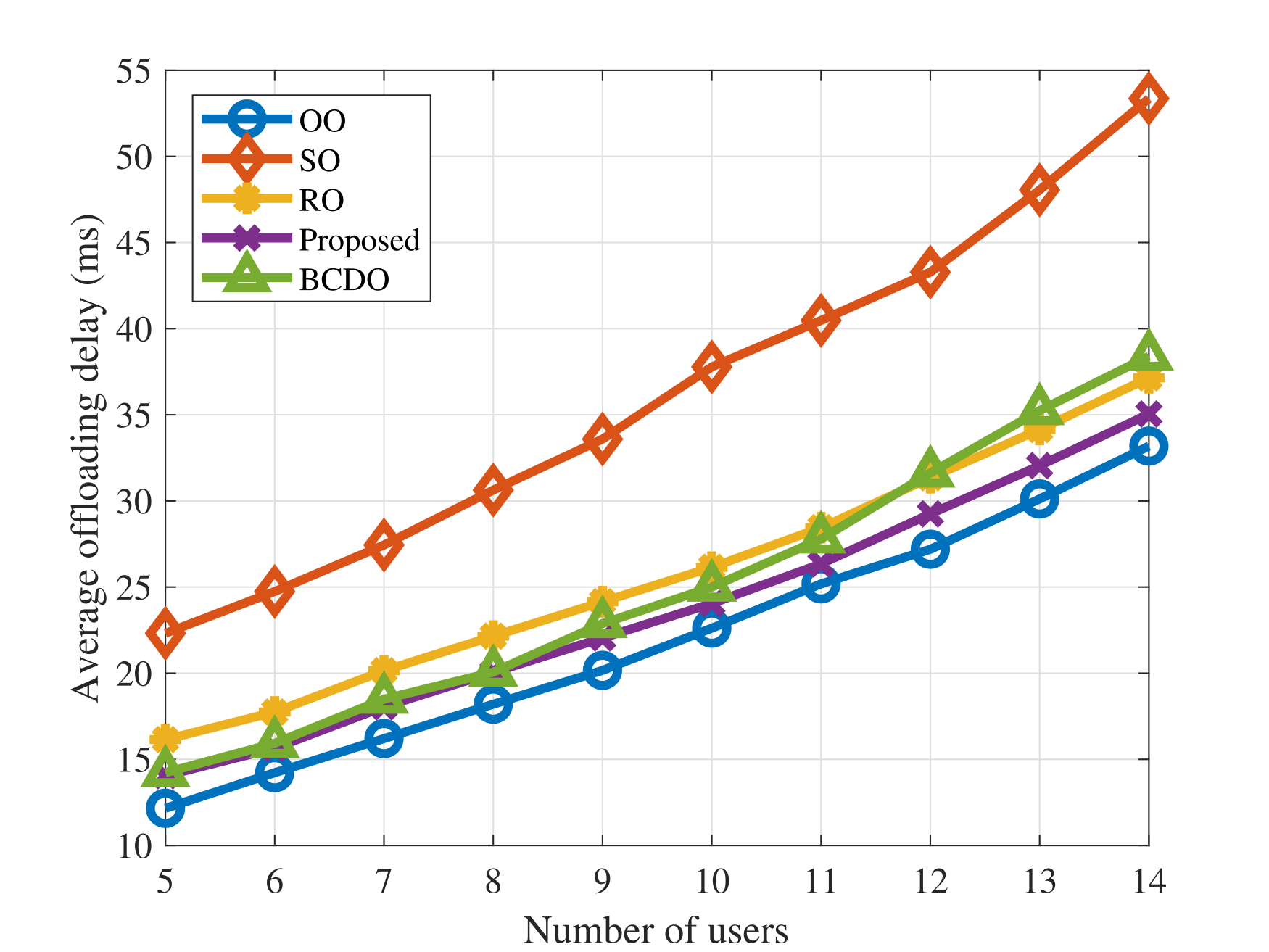}}
    \label{1a}
	  \subfloat[Consensus delay under different user numbers.]{
        \includegraphics[width=0.3\linewidth]{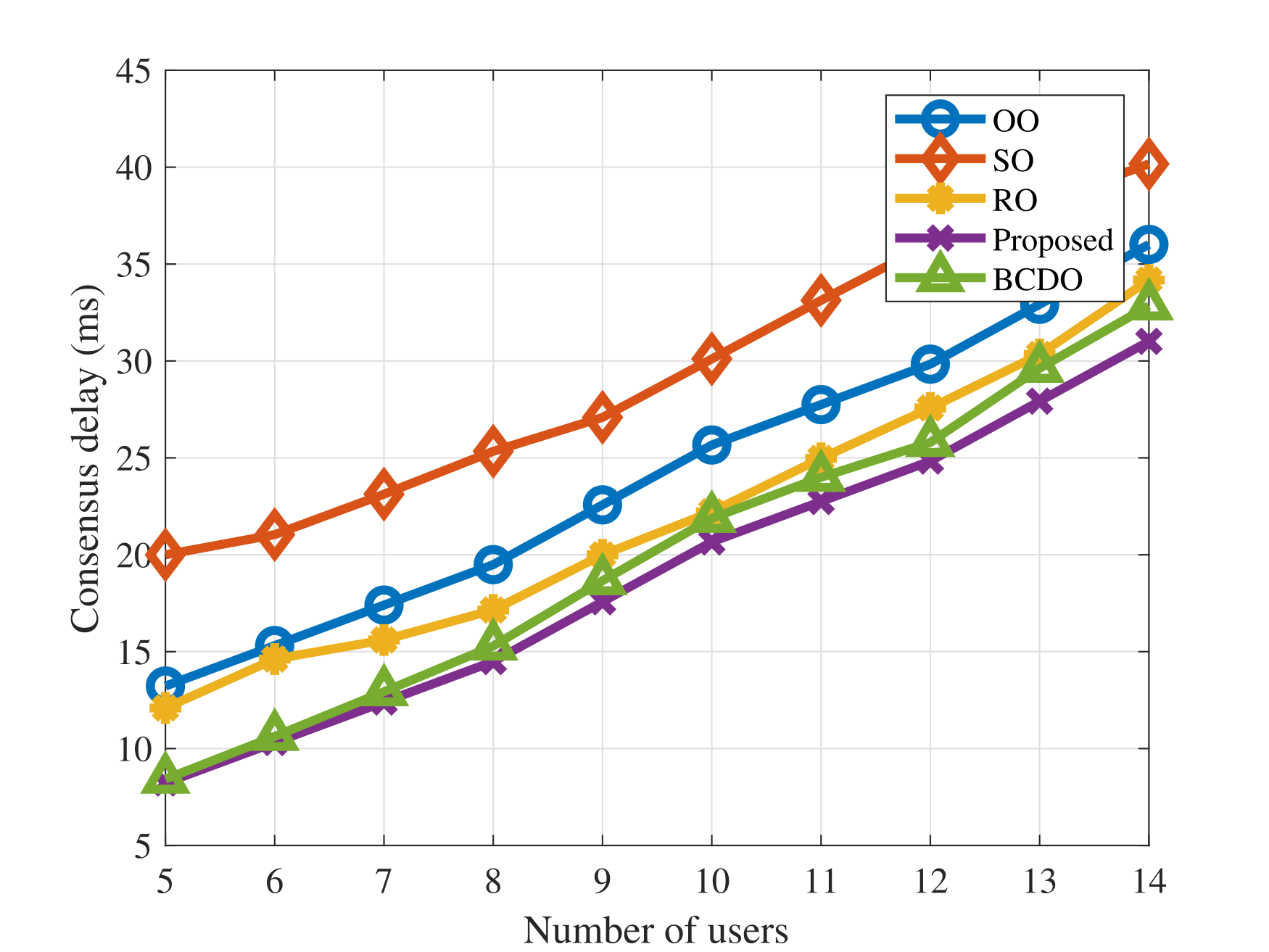}}
    \label{1b}
         \subfloat[Total delay under different user numbers.]{
       \includegraphics[width=0.3\linewidth]{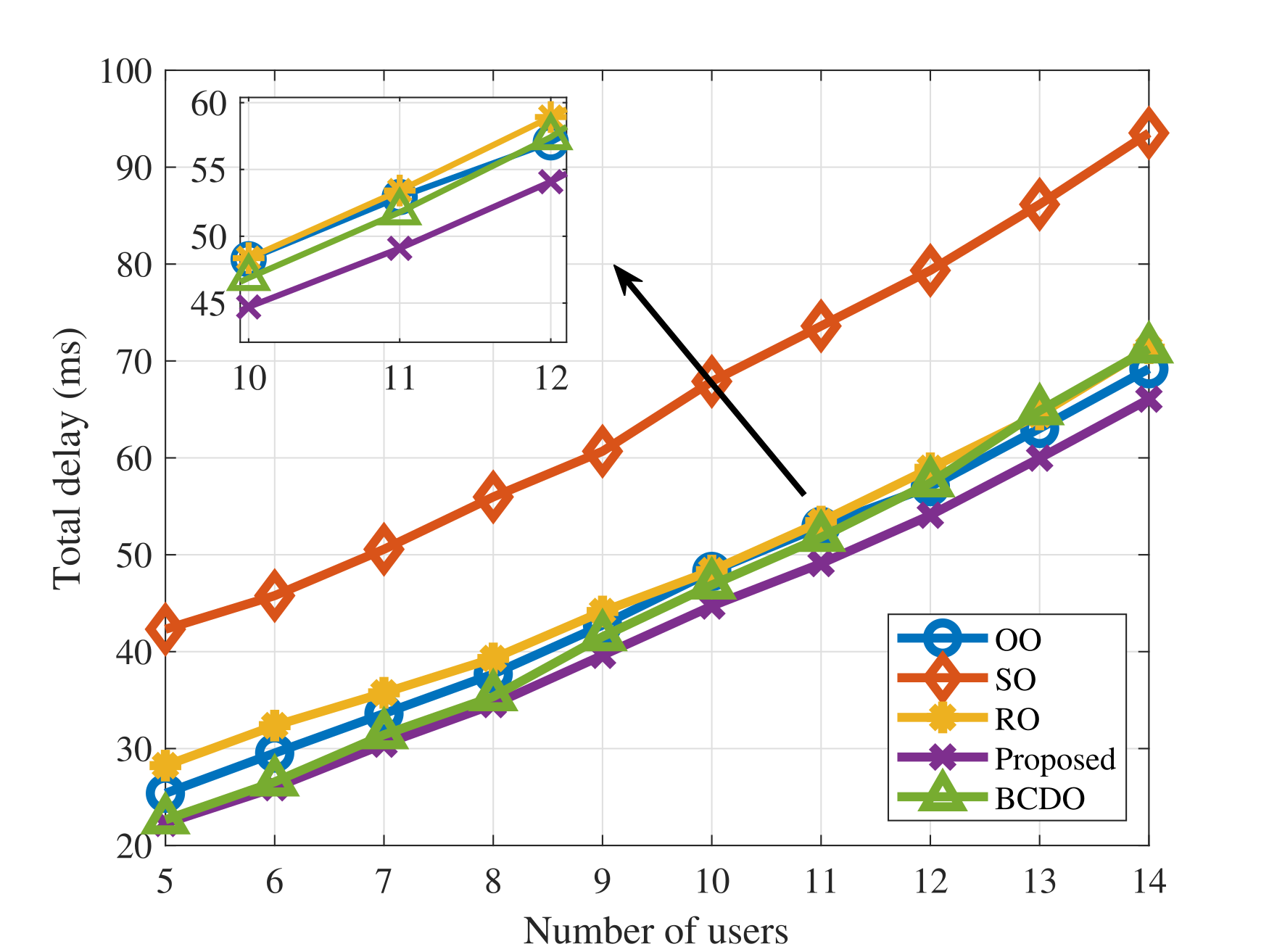}}
    \label{1c}
        \caption{Delay performance under different user numbers.}
	\end{figure*}
	
	\begin{figure*}[tbp]
    \centering
    \captionsetup{justification=centering}
	  \subfloat[Offloading energy consumption under different user numbers.]{
        \includegraphics[width=0.3\linewidth]{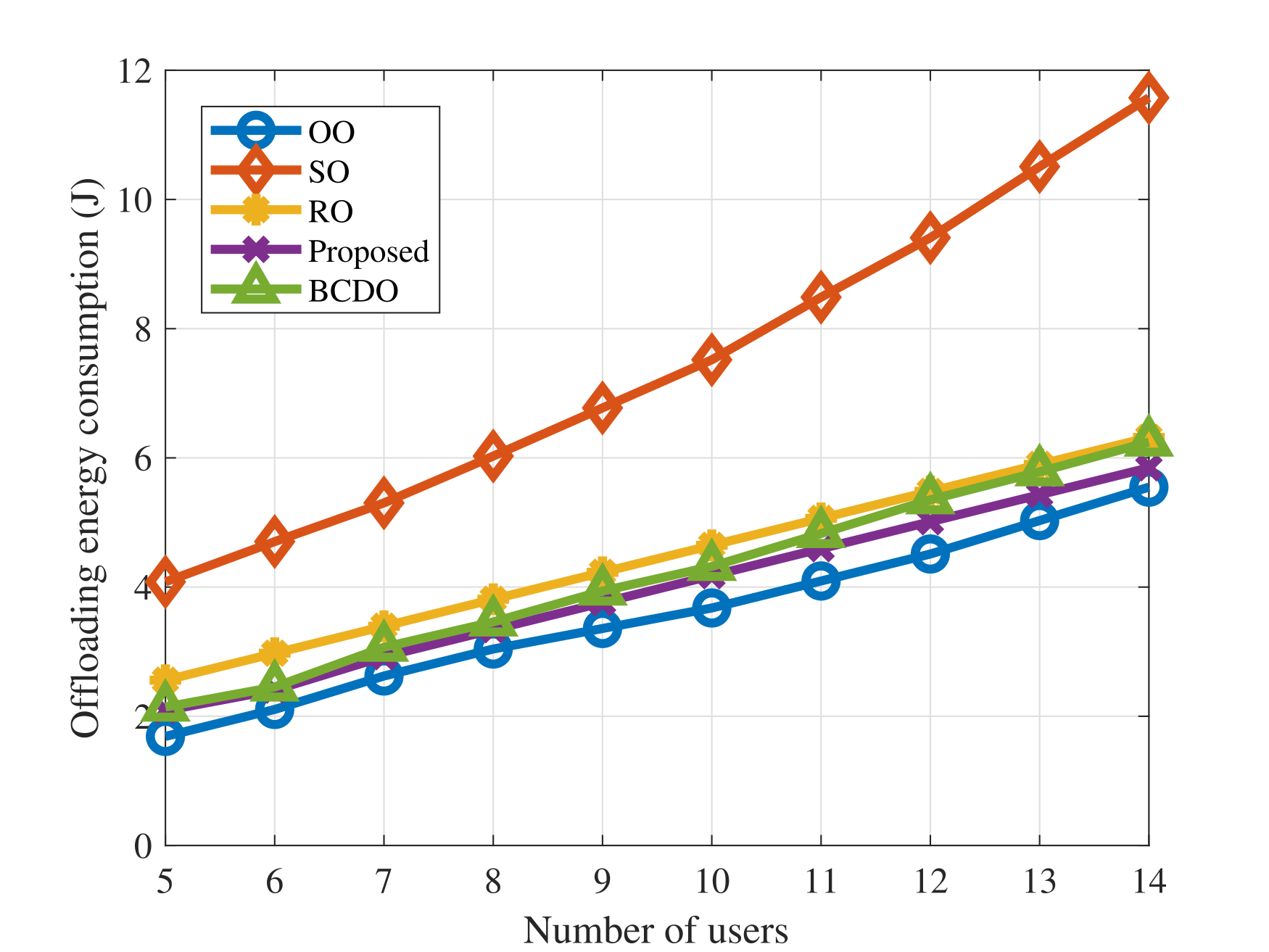}}
    \label{1d}
	  \subfloat[Consensus energy consumption under different user numbers.]{
        \includegraphics[width=0.3\linewidth]{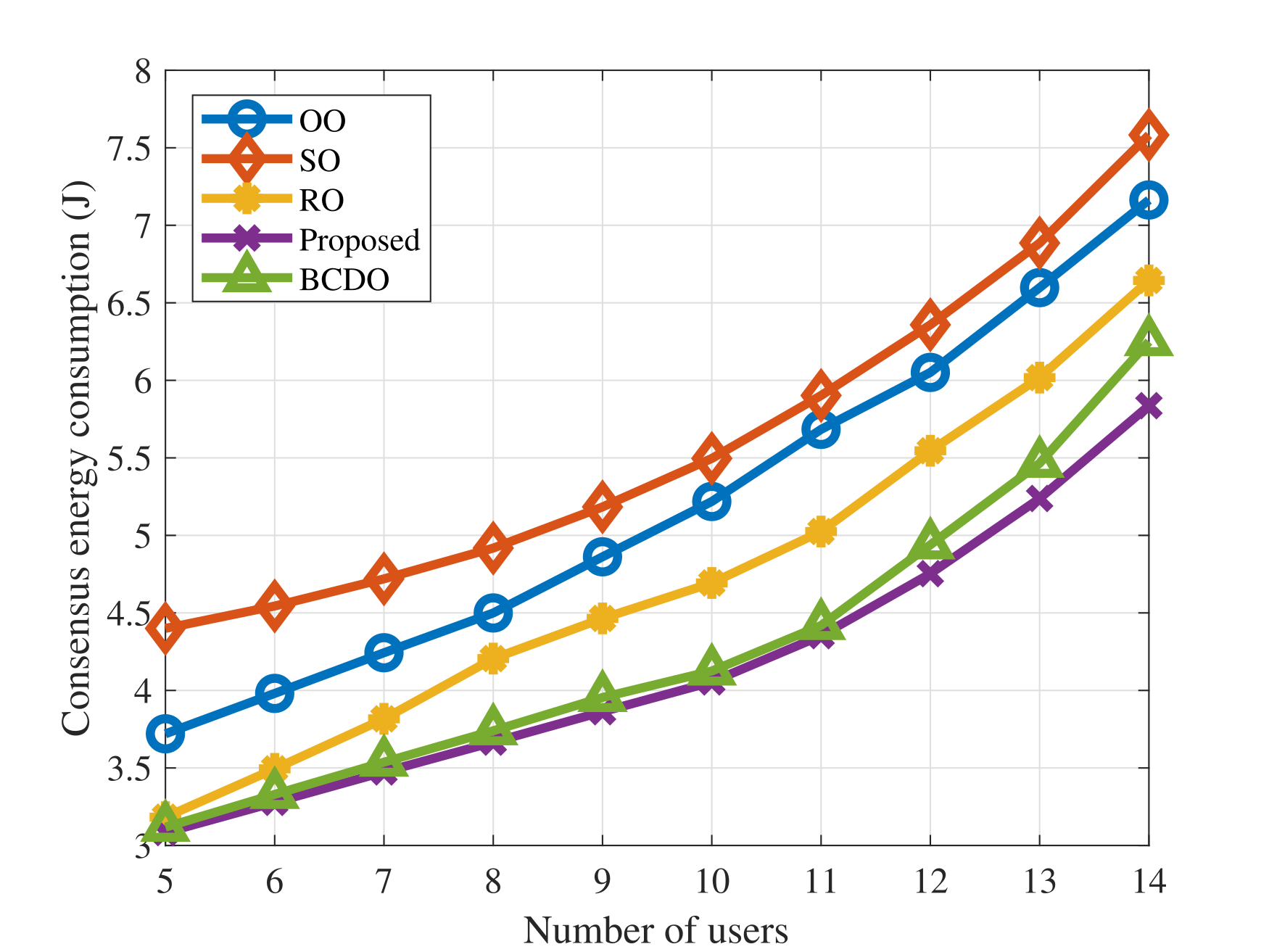}}
    \label{1e}
	  \subfloat[Total energy consumption under different user numbers.]{
       \includegraphics[width=0.3\linewidth]{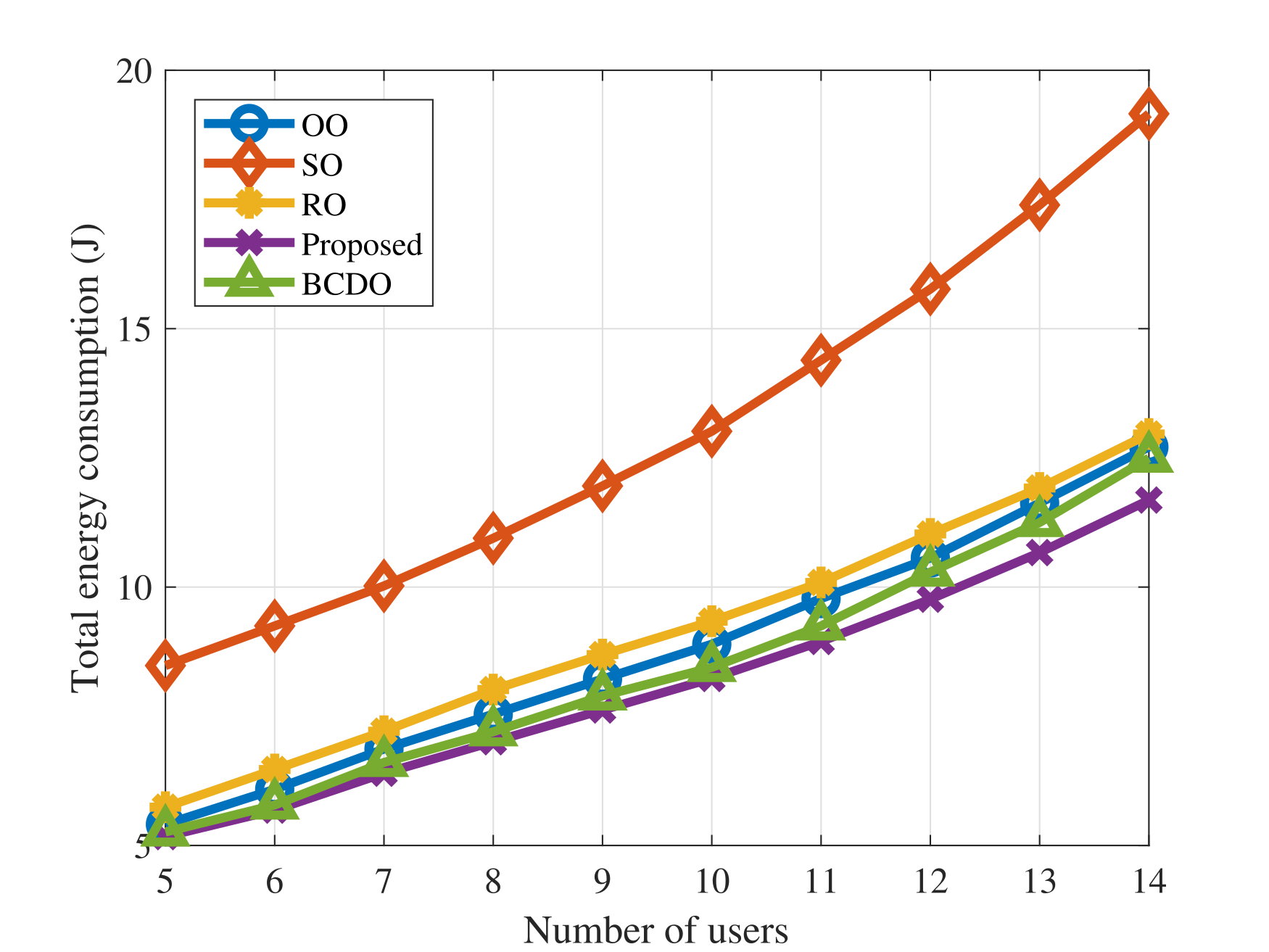}}
    \label{1f}
	  	  \caption{Energy consumption under different user numbers.}
	\end{figure*}

\subsubsection{Delay Performance}
\par Next, we analyse the delay performance when the number of users changes. We compare the average offloading delay $D^{o} = \frac{1}{N}\sum_{n \in \mathcal{N}}D_n^{o}$, the consensus delay $D^{c}$ and the total delay $D^{a} = D^{o} + D^{c}$, respectively.

\par  It can be observed from Fig. 5 that when the number of users increases, the total delay, the average offloading delay of users, and the consensus delay will also increase. This is because the increase in offloading tasks causes resource-constrained APs to spend more time processing them. Hence, the offloading delay will increase. Similarly, since APs spend more resources in the offloading process, the available computing resources and bandwidth in the consensus process will reduce, resulting the increase of consensus delay. Moreover, the total delay in \textbf{SO} is significantly higher than that in the other three schemes. The reason is that although transmission delay will reduce when users select the closest AP, the service quality is not optimal. In contrast, by introducing a cooperative APs clustering mechanism, users can be served by multiple APs, thereby significantly improving the efficiency of resource utilization and reducing the total delay.

\par As we can see, \textbf{OO} can achieve the lowest offloading delay. However, it does not consider the consensus delay and focuses too much on task offloading, making the available resources in the consensus process insufficient, resulting in a relatively large total delay. In \textbf{RO}, the optimization of task offloading and consensus is jointly considered, thus the offloading delay is less than that in \textbf{SO}, and the consensus delay is also less than that in \textbf{SO} and \textbf{OO}. However, RAFT does not consider the resource usage of APs when selecting \emph{Leader}, thereby election conflict is prone to happen, resulting a relatively large consensus delay.

\par {\color{black}In contrast, our proposed scheme uses an improved resource-aware RAFT consensus mechanism while jointly considering the optimization of task offloading and consensus. When $N=14$, the offloading delay of our proposed scheme is only $5.67\%$ more than that of \textbf{OO}, but $33.96\%$ less than that of \textbf{SO}, which can still be maintained within $35$ ms. Consensus delay of our proposed scheme is the smallest of the four schemes, which can keep at $31.01$ ms, $11.43\%$ less than that of \textbf{SO}. The total delay is also the smallest compared to reference schemes, which is about $66$ ms and is $21.4\%$ less than the total delay of \textbf{SO}. In addition, \textbf{BCDO} is basically consistent with the proposed scheme when the number of users is small, but when the number of users increases, the solution space will become larger and the performance of \textbf{BCDO} will deteriorate. Therefore, it can be concluded that our scheme can be applied to delay-sensitive applications and maintain a low-latency performance.}

\begin{figure*}[t]
    \centering
    \captionsetup{justification=centering}
	  \subfloat[Impact of size of block on energy consumption.]{
        \includegraphics[width=0.47\linewidth]{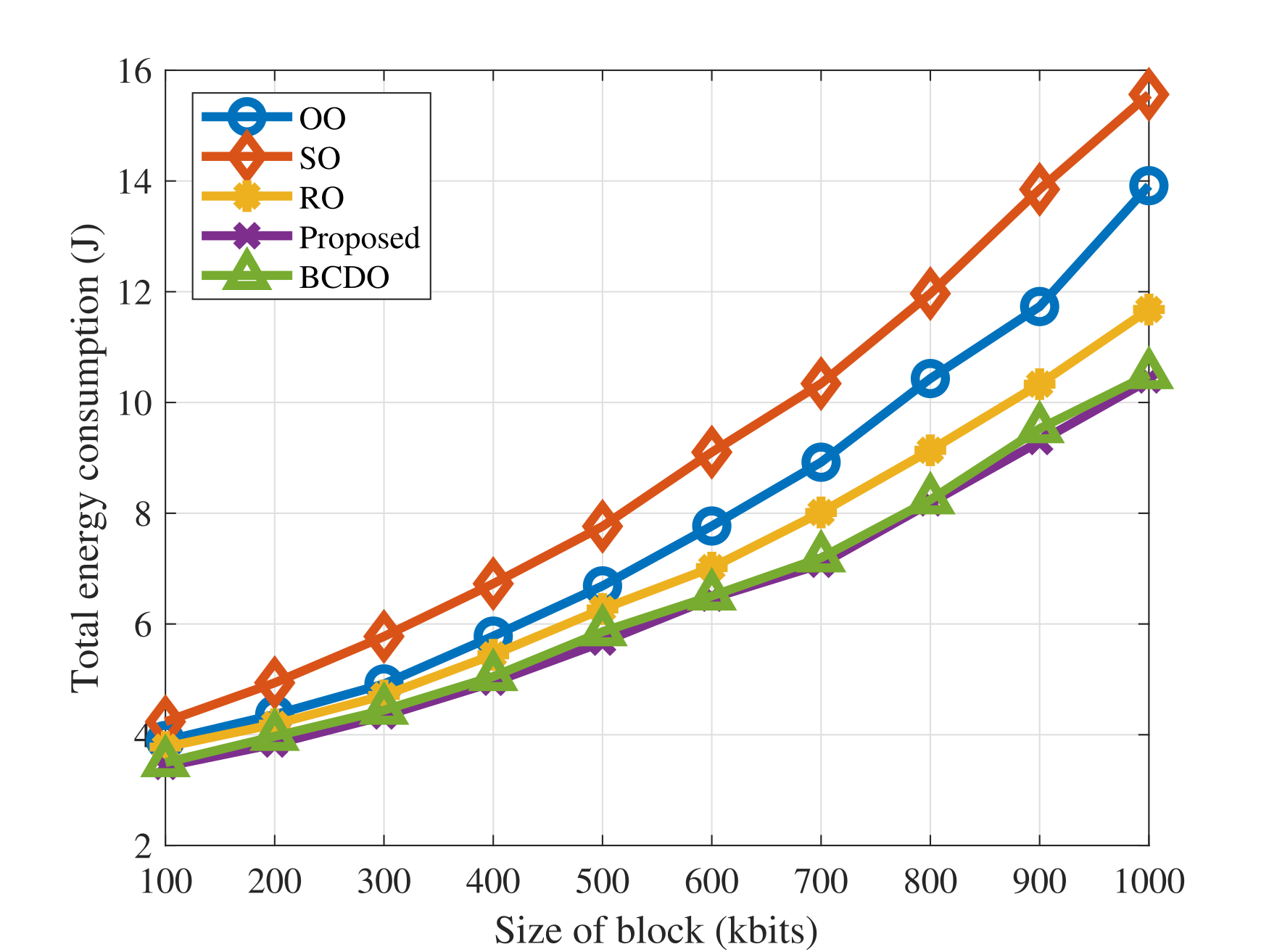}}
    \label{1a}
	  \subfloat[Impact of number of APs on energy consumption.]{
        \includegraphics[width=0.47\linewidth]{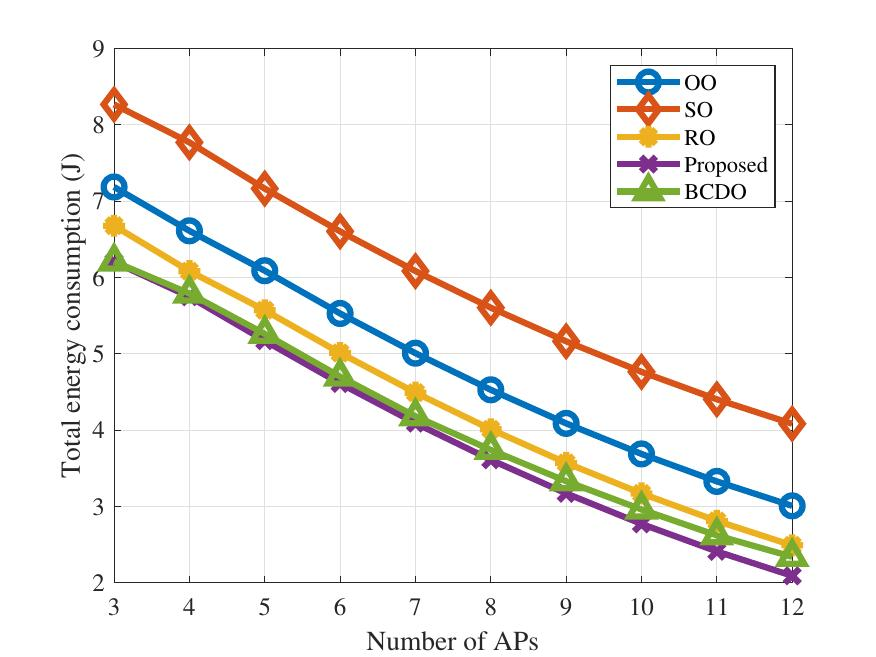}}
    \label{1b}
	  \caption{Total energy consumption under different parameters}
	  \label{fig1} 
	\end{figure*}

\subsubsection{Energy Consumption Performance}
In this section, we evaluate the impact of different parameters on the performance of the offloading energy consumption $E^{o} = \sum_{n \in \mathcal{N}}{E_n^{o}} + \sum_{m \in \mathcal{M}}{E_m^{o}}$,  the consensus energy consumption $E^{c} =\sum_{m \in \mathcal{M}}{E_m^{c}} $ and the total energy consumption $E^{a}$.

As shown in Fig. 6, when the number of users increases, the energy consumption in offloading and consensus will increase. The reason is that energy consumption and delay are positively correlated, when the offloading and the consensus delay increase, the corresponding energy consumption will increase too. {\color{black}When the number of users $N=14$, the offloading energy consumption of our proposed scheme is only $5.42\%$ more than that of \textbf{OO}, but $49.52\%$ less than that of \textbf{OO}, consuming $5.84$ J of energy when $N=14$. Our proposed scheme can not only consume the least energy in the consensus, $25.54\%$ less than that of \textbf{SO}, consume $5.83$ J of energy, but also consume the least energy in total, $39.01\%$ less than that of \textbf{SO}, consuming $11.68$ J of energy in total.}

\par Fig. 7 (a) shows the energy consumption under different block sizes. When the block size increases, energy consumption of the system will increase. Because the increase in the delay for APs to generate blocks during the consensus will consume more energy. In \textbf{SO}, \textbf{OO}, and \textbf{RO}, the total energy consumption increase significantly with the increase in the block size, however, in the proposed scheme, energy consumption increases slowly since the delay-based RAFT mechanism can optimally select APs with more resources as the \emph{Leader}, thereby minimizing the delay and the energy consumption for communication and block generation. {\color{black} Our scheme maintains the lowest energy consumption when the size of the block increase, which is $33.16\%$ less than that of \textbf{SO}, consuming $10.4$ J of energy when block size is $1000$ kbits.}

{\color{black}

\par Fig. 7 (b) depicts the impact of number of APs on the total energy consumption. When the number of APs increases, the total energy consumption will decrease, which is consistent with the conclusion in Fig. 3 (c). When the number of APs increases, the computing resources and bandwidth allocated to users will be more abundant, but more energy will be consumed in the process of task processing and consensus, and the gain will be gradually offset. When number of APs $M=12$, the total energy consumption of the proposed scheme is $48.77\%$ less than that of \textbf{SO}, consuming $2.09$ J of energy.

}

\section{Conclusion and Future Work}
In this paper, we propose an energy-efficient blockchain-enabled UC-MEC framework. In the proposed system, each user will be provided with wireless transmission and task offloading services by a specific AP cluster. After the task processing is completed, all APs will run the consensus mechanism to record the resource transaction transactions to the blockchain to ensure the security of the network. A resource-aware RAFT consensus mechanism is proposed which considers the computing resources and bandwidth consumed by the data transmission and block duplication. To minimize the total energy consumption, we jointly optimize the clustering and the resource allocation strategy in the task offloading and the consensus process. To solve the problem, we decompose it into $M$ sub-problems based on ADMM so that each AP can solve the sub-problem in parallel. Simulation results show that our proposed scheme has high effectiveness under different parameters and can significantly reduce the total energy consumption when compared with traditional MEC.

\bibliographystyle{IEEEtran}
\bibliography{IEEEabrv,ref}

\end{document}